\documentclass[a4paper,12pt]{article}
\usepackage{amssymb}
\usepackage{amsmath}
\usepackage{epsfig}
\usepackage{subfig}
\usepackage{caption}
\usepackage{graphicx}
\usepackage{tikz}
\usepackage{latexsym}
\usepackage{rotating}
\usepackage{titlesec}
\newcommand{\squeezeup}{\vspace{-10mm}}
\usetikzlibrary{matrix}

\usepackage{relsize}
\title{Local and nonlocal $(2+1)$-dimensional Maccari systems and their soliton solutions}

\author{Asl{\i} Pekcan \thanks{aslipekcan@hacettepe.edu.tr} \\
{\small Department of Mathematics, Faculty of Science} \\
{\small Hacettepe University, 06800 Ankara - Turkey}
}

\date{\nonumber}
\begin{document}
\maketitle
\date{\nonumber}
\begin{abstract}
In this work, by using the Hirota bilinear method, we obtain one- and two-soliton solutions of integrable $(2+1)$-dimensional $3$-component Maccari system which is used as a model describing isolated waves localized in a very small part of space and related to very well-known systems like nonlinear Schr\"{o}dinger, Fokas, and long wave resonance systems. We represent all local and Ablowitz-Musslimani type nonlocal reductions of this system and obtain new integrable systems. By the help of reduction formulas and soliton solutions of the $3$-component Maccari system, we obtain one- and two-soliton solutions of these new integrable local and nonlocal reduced $2$-component Maccari systems. We also illustrate our solutions by plotting their graphs for particular values of the parameters.\\

\end{abstract}

\section{Introduction}

The generalized $(2+1)$-dimensional $(N+1)$-component Maccari system is given by
\begin{align}
&iu_{k,t}+u_{k,xx}+pu_k=0, \quad k=1,2,\ldots,N,\label{eqn1}\\
&p_y=\sum_{k=1}^N(\sigma_k u_k\bar{u}_k)_x,\label{eqn2}
\end{align}
where $p=p(x,y,t)$, $\sigma_k=\pm 1$, $u_k=u_k(x,y,t)$ for $k=1,2,\ldots,N$, and the bar notation is used for complex conjugation. Here $u_k$ denote $N$ different complex short wave amplitudes and $p$
denotes the real long wave amplitude. The system for $N=2$ is
\begin{align}
&iu_t+u_{xx}+pu=0,\label{prN=2-a}\\
&iv_t+v_{xx}+pv=0,\label{prN=2-b}\\
&p_y=(\sigma_1u\bar{u}+\sigma_2v\bar{v})_x,\label{prN=2-c}
\end{align}
where $\sigma_k=\pm 1$, $k=1, 2$. This $(2+1)$-dimensional $3$-component system was first derived by Maccari \cite{Maccari1997} from Nizhnik-Novikov-Veselov equation which is an S-integrable equation, with a reduction technique based on Fourier decomposition and space-time rescalings.  In \cite{Maccari2020}, it was noted that Maccari system can also be obtained from many nonlinear partial differential equations by following the same technique. This method preserves the integrability that is the Maccari system (\ref{prN=2-a})-(\ref{prN=2-c}) is also integrable. Painlev\'{e} analysis was used to investigate the integrability of the Maccari system by Uthayakumar et al. in \cite{Uthayakumar}.

The Maccari system (\ref{prN=2-a})-(\ref{prN=2-c}) is related to some well-known equations. If $y=x$ it reduces to coupled nonlinear Schr\"{o}dinger (NLS) equation \cite{Zakharov}. When $y=t$, we have the coupled long-wave resonance system \cite{Craik}. We get $(2+1)$-dimensional extension of NLS equation introduced by Fokas \cite{Fokas} when $u=\bar{v}$.

Maccari system has many applications in hydrodynamics, plasma physics, Bose-Einstein condensates, and nonlinear optics. It is a model describing isolated waves which are localized in a very small part of space. There are several works on obtaining solutions of the Maccari system (\ref{prN=2-a})-(\ref{prN=2-c}). By KP hierarchy reduction method, bright-dark mixed soliton solutions \cite{HanChen1}, \cite{HanChen2}, multi-dark soliton solutions \cite{Xu}, semi-rational solutions \cite{Wazwaz} of the Maccari system were obtained. Solitoff and dromion solutions \cite{HZZ}, solutions in terms of Jacobi elliptic functions \cite{HLM}, folded solitary waves and foldons \cite{HuangZhang} were found by using variable separation approach. Half bright, bright, dark, half dark, and combined solitons were given by extended modified auxiliary equation mapping technique in \cite{CCS}. By the Hirota bilinear method, soliton solutions \cite{Uthayakumar}, lump and rogue wave solutions \cite{Yuan}, and dromions \cite{Zhang} were derived. By means of the B\"{a}cklund transformation, peakon and compacton solutions of the Maccari system were obtained in \cite{ZMH}.

Recently, there is a huge interest on constructing integrable nonlocal equations and finding various types of solutions to these equations.
The nonlocal reductions were first introduced by Ablowitz and Musslimani \cite{AbMu1}-\cite{AbMu3}. It has been indicated that systems admitting nonlocal reductions have discrete symmetry transformations that leave the systems invariant. In \cite{origin} we showed that a special case of discrete symmetry transformations gives the nonlocal reductions of the same systems. As examples of $(1+1)$-dimensional nonlocal integrable equations we can mention some famous equations; nonlocal NLS \cite{AbMu1}-\cite{AbMu3}, \cite{chen}-\cite{Ma1}, modified Korteweg-de Vries (mKdV) \cite{AbMu2}-\cite{chen}, \cite{GurPek3}, \cite{GurPek2}-\cite{Yan}, Fordy-Kulish \cite{Gurses}, hydrodynamic type \cite{GurPekKos}, KdV \cite{GurPek6} equations, and so on. For $(2+1)$-dimensional nonlocal integrable equations one can check the references \cite{RZFH}-\cite{Rao} for Davey-Stewartson equations, \cite {Mihalache} for Fokas equations, \cite{XZhu} and \cite{Guo} for $(2+1)$-dimensional mKdV equations, \cite{LiuLi} for $(2+1)$-dimensional NLS equations, \cite{ZhangChen} for Kadomtsev-Petviashvili equations, \cite{GurPek4} and \cite{GurPek5} for $(2+1)$-dimensional negative AKNS equations, and so on. Possible applications of nonlocal equations have been discussed in \cite{Lou1}-\cite{magnet}.

As far as we know there is no work about the local and nonlocal reductions of the integrable $(2+1)$-dimensional $3$-component Maccari system
even there are many works about different type of solutions of this system in the literature. We note that if the nonlocal (or local) reductions of integrable systems are done consistently then the reduced equations are also integrable. Finding new integrable local and nonlocal systems is very important in the integrable systems and soliton theory since they have rich physical structures and features leading to many developments in describing nature. In this work we obtain new integrable local and nonlocal reduced systems from the $(2+1)$-dimensional $3$-component Maccari system.

To be able to analyze all local and nonlocal reductions of the Maccari system we first let $\displaystyle t\rightarrow \frac{a}{i}t$ and consider the following $(2+1)$-dimensional $3$-component Maccari system:
\begin{align}
&au_t+u_{xx}+pu=0,\label{N=2-a}\\
&av_t+v_{xx}+pv=0,\label{N=2-b}\\
&p_y=(\sigma_1u\bar{u}+\sigma_2v\bar{v})_x,\label{N=2-c}
\end{align}
where $a$ is an arbitrary constant. For this system we have two different local reductions,
\begin{eqnarray}
&i)\, v(x,y,t)= \rho u(x,y,t), \label{non0}\\
&ii)\, v(x,y,t)=\rho \bar{u}(x,y,t), \label{non00}
\end{eqnarray}
where $\rho$ is a real constant and $\bar{u}$ is the complex conjugate of the function $u$. Both of these reductions give the same $(2+1)$-dimensional $2$-component Maccari system,
\begin{align}
&au_t+u_{xx}+pu=0,\\
&p_y=(\sigma_1+\sigma_2\rho^2)(u\bar{u})_x.
\end{align}
The Maccari system (\ref{N=2-a})-(\ref{N=2-c}) has two different consistent nonlocal reductions. The first one of these reductions is
\begin{equation}
i)\, v(x,y,t)=\rho u(\varepsilon_1x,\varepsilon_2y,\varepsilon_3t),
\end{equation}
where $\rho$ is a real constant and $\varepsilon_j^2=1$, $j=1,2,3$ with the constraints $\varepsilon_3=1$ and $\varepsilon_1\varepsilon_2\sigma_1=\sigma_2\rho^2$. This reduction gives the following integrable space reversal nonlocal $(2+1)$-dimensional $2$-component Maccari systems:
\begin{align}
&au_t(x,y,t)+u_{xx}(x,y,t)+p(x,y,t)u(x,y,t)=0,\\
&p_y(x,y,t)=(\sigma_1u(x,y,t)\bar{u}(x,y,t)+\sigma_2u(\varepsilon_1x,\varepsilon_2y,t)\bar{u}(\varepsilon_1x,\varepsilon_2y,t))_x.
\end{align}
The second nonlocal reduction of the Maccari system is given by
\begin{equation}
ii)\, v(x,y,t)=\rho \bar{u}(\varepsilon_1x,\varepsilon_2y,\varepsilon_3t),
\end{equation}
yielding seven different nonlocal systems which are nonlocal time (T-), space (S-), and space-time (ST-) reversal Maccari systems represented by
\begin{align}
&au_t(x,y,t)+u_{xx}(x,y,t)+p(x,y,t)u(x,y,t)=0,\\
&p_y(x,y,t)=(\sigma_1u(x,y,t)\bar{u}(x,y,t)+\sigma_2u(\varepsilon_1x,\varepsilon_2y,\varepsilon_3t)
\bar{u}(\varepsilon_1x,\varepsilon_2y,\varepsilon_3t))_x,
\end{align}
where $a=\bar{a}\varepsilon_3$ and $\varepsilon_1\varepsilon_2\sigma_1=\sigma_2\rho^2$.

In this work we study soliton solutions of local and nonlocal reduced $(2+1)$-dimensional $2$-component Maccari systems. We first obtain one- and two-soliton solutions of the $3$-component Maccari system by Hirota bilinear method \cite{Hirota1}-\cite{Yu3} which has been used as a powerful and practical tool to obtain multi-soliton solutions of integrable systems analytically. Then by using local and nonlocal Ablowitz-Musslimani type reductions, we obtain soliton solutions of the integrable local and nonlocal reduced $2$-component Maccari systems. We present the graphs of some solutions for particular values of the parameters.

The lay out of the paper is as follows. In Section 2, we obtain one- and two-soliton solutions of the $3$-component Maccari system by Hirota bilinear method. Then we present local and nonlocal reductions of this system in Section 3 and 4, respectively. In Section 5, we derive soliton solutions of integrable local and nonlocal reduced $2$-component Maccari systems. We also give some examples of one- and two-soliton solutions of reduced Maccari systems.

\section{Hirota bilinear method for $3$-component Maccari system}

Here by using the Hirota bilinear method \cite{Hirota1}-\cite{Hietarinta} we will obtain one- and two-soliton solutions of $3$-component Maccari system (\ref{N=2-a})-(\ref{N=2-c}). Let
\begin{equation}\displaystyle
u=\frac{g}{f},\quad v=\frac{h}{f},\quad p=2(\ln f)_{xx}.
\end{equation}
Here $g(x,y,t)$, $h(x,y,t)$ are complex-valued functions and $f(x,y,t)$ is a real-valued function. We obtain the Hirota bilinear form of the system (\ref{N=2-a})-(\ref{N=2-c}) as
\begin{align}
&(aD_t+D_x^2)\{g\cdot f\}=0,\label{N=2Hirota-a}\\
&(aD_t+D_x^2)\{h\cdot f\}=0,\label{N=2Hirota-b}\\
&D_xD_y\{f\cdot f\}=\sigma_1g\bar{g}+\sigma_2h\bar{h}.\label{N=2Hirota-c}
\end{align}
Here $D$ is a special differential operator called Hirota $D$-operator given by{\small
\begin{equation}
D_t^nD_x^mD_y^r\{F\cdot G\}=\Big(\frac{\partial}{\partial t}-\frac{\partial}{\partial t'}\Big)^n\Big(\frac{\partial}{\partial x}-\frac{\partial}{\partial x'}\Big)^m\Big(\frac{\partial}{\partial y}-\frac{\partial}{\partial y'}\Big)^r\,F(x,y,t)G(x',y',t')|_{x'=x,t'=t,y'=y}.
\end{equation}}
 We expand the functions $g$, $h$, and $f$ in the form of power series as
 \begin{align}
    & g=\epsilon g_1+\epsilon^3 g_3+\epsilon^5 g_5+\cdots, \nonumber\\
    & h=\epsilon h_1+\epsilon^3 h_3+\epsilon^5 h_5+\cdots, \nonumber \\
    & f=1+ \epsilon^2 f_2+\epsilon^4 f_4+\cdots.\label{generalexpansion}
 \end{align}

 We substitute the above expansion into the Hirota bilinear form (\ref{N=2Hirota-a})-(\ref{N=2Hirota-c}) and make the coefficients of $\epsilon^m$, $m=1,2,\ldots$ to vanish. Note that choosing the functions $g_1$ and $h_1$ as exponential functions determines the other functions $g_j, h_j,$ and $f_j$ as exponential functions. Due to this fact, when we consider $N$ soliton solution we have $g_{2j+1}=h_{2j+1}=f_{2j+2}=0$ for $j\geq N$. In the following two sections, one- and two-soliton solutions of $(2+1)$-dimensional $3$-component Maccari system will be given.

\subsection{One-soliton solutions of $3$-component Maccari system}

In order to construct one-soliton solution, we take
\begin{equation}
g_1=e^{\theta_1},\quad h_1=e^{\theta_2},
\end{equation}
in the expansion (\ref{generalexpansion}) and make the coefficients of $\epsilon^m$, $m=1,2,\ldots$ to vanish in (\ref{N=2Hirota-a})-(\ref{N=2Hirota-c}).
Here $\theta_j=k_jx+l_jy+\omega_jt+\delta_j$, $j=1,2$. By the above choice of $g_1$ and $h_1$ we have $g_{2j+1}=h_{2j+1}=f_{2j+2}=0$ for $j\geq 1$.
From the coefficients of $\epsilon$, we get the dispersion relations,
\begin{equation}\label{dispersion}
\omega_j=-\frac{k_j^2}{a},\, j=1,2.
\end{equation}
The coefficient of $\epsilon^2$ yields the function $f_2$ as
\begin{equation}
f_2=\frac{\sigma_1e^{\theta_1+\bar{\theta}_1}}{2(k_1+\bar{k}_1)(l_1+\bar{l}_1)}+\frac{\sigma_2e^{\theta_2+\bar{\theta}_2}}{2(k_2+\bar{k}_2)(l_2+\bar{l}_2)}.
\end{equation}
The coefficient of $\epsilon^4$
\begin{equation}
f_{2,xy}f_2-f_{2,x}f_{2,y}=0
\end{equation}
is satisfied if
\begin{equation}
(l_1+\bar{l}_1-l_2-\bar{l}_2)(k_1+\bar{k}_1-k_2-\bar{k}_2)=0.
\end{equation}
Therefore we have two possibilities:
\begin{align}
&1)\, l_2+\bar{l}_2=l_1+\bar{l}_1,\label{pos1}\\
&2)\, k_2+\bar{k}_2=k_1+\bar{k}_1.\label{pos2}
\end{align}
The coefficients of $\epsilon^3$ are
\begin{align}
&a(g_{1,t}f_2-g_1f_{2,t})+g_{1,xx}f_2-2g_{1,x}f_{2,x}+g_1f_{2,xx}=0,\\
&a(h_{1,t}f_2-h_1f_{2,t})+h_{1,xx}f_2-2h_{1,x}f_{2,x}+h_1f_{2,xx}=0.
\end{align}
To satisfy the above equations we must have
\begin{equation}\label{onesolcond}
\bar{a}=-a, \quad k_1=k_2,
\end{equation}
and so that (\ref{pos2}) holds. Without loss of generality take $\epsilon=1$. Hence one-soliton solution of the $3$-component Maccari system (\ref{N=2-a})-(\ref{N=2-c}) for $\bar{a}=-a$ is given by $(u(x,y,t),v(x,y,t),p(x,y,t))$ where
\begin{align}\displaystyle
&u(x,y,t)=\frac{e^{k_1x+\omega_1t+l_1y+\delta_1}}{1+\frac{e^{(k_1+\bar{k}_1)x+(\omega_1+\bar{\omega}_1)t}}{2(k_1+\bar{k}_1)}
[\frac{\sigma_1e^{(l_1+\bar{l}_1)y+\delta_1+\bar{\delta}_1}}{(l_1+\bar{l}_1)}+\frac{\sigma_2e^{(l_2+\bar{l}_2)y+\delta_2+\bar{\delta}_2}}{(l_2+\bar{l}_2)}       ]},\\
&v(x,y,t)=\frac{e^{k_1x+\omega_1t+l_2y+\delta_2}}{1+\frac{e^{(k_1+\bar{k}_1)x+(\omega_1+\bar{\omega}_1)t}}{2(k_1+\bar{k}_1)}
[\frac{\sigma_1e^{(l_1+\bar{l}_1)y+\delta_1+\bar{\delta}_1}}{(l_1+\bar{l}_1)}+\frac{\sigma_2e^{(l_2+\bar{l}_2)y+\delta_2+\bar{\delta}_2}}{(l_2+\bar{l}_2)}       ]},
\end{align}
and
\begin{equation}\label{p(x,y,t)}
p(x,y,t)=\frac{(k_1+\bar{k}_1)e^{(k_1+\bar{k}_1)x+(\omega_1+\bar{\omega}_1)t}[\frac{\sigma_1e^{(l_1+\bar{l}_1)y+\delta_1
+\bar{\delta}_1}}{(l_1+\bar{l}_1)}+\frac{\sigma_2e^{(l_2+\bar{l}_2)y+\delta_2+\bar{\delta}_2}}{(l_2+\bar{l}_2)}]   }{(1+\frac{e^{(k_1+\bar{k}_1)x+(\omega_1+\bar{\omega}_1)t}}{2(k_1+\bar{k}_1)}
[\frac{\sigma_1e^{(l_1+\bar{l}_1)y+\delta_1+\bar{\delta}_1}}{(l_1+\bar{l}_1)}+\frac{\sigma_2e^{(l_2+\bar{l}_2)y+\delta_2+\bar{\delta}_2}}{(l_2+\bar{l}_2)}       ])^2},
\end{equation}
where the dispersion relations (\ref{dispersion}) and the conditions (\ref{onesolcond}) hold. Here $k_1, l_1, l_2, \delta_1, \delta_2$ are some arbitrary complex numbers.

\subsection{Two-soliton solutions of $3$-component Maccari system}

To construct two-soliton solution, we take
\begin{equation}\label{g_1h_1}
g_1=e^{\theta_1}+e^{\theta_2},\quad h_1=e^{\eta_1}+e^{\eta_2},
\end{equation}
in the expansion (\ref{generalexpansion}), where $\theta_j=k_jx+l_jy+\omega_jt+\delta_i$, $\eta_j=r_jx+m_jt+s_jy+\alpha_j$ for $j=1, 2$.
We make the coefficients of $\epsilon^m$, $m=1,2,\ldots$ to vanish in (\ref{N=2Hirota-a})-(\ref{N=2Hirota-c}).
By the choice of $g_1$ and $h_1$ we have $g_{2j+1}=h_{2j+1}=f_{2j+2}=0$ for $j\geq 2$.
From the coefficients of $\epsilon$, we get the dispersion relations,
\begin{equation}\label{dispersiontwosoliton}
\omega_j=-\frac{k_j^2}{a},\, m_j=-\frac{r_j^2}{a},\, j=1, 2.
\end{equation}
The coefficient of $\epsilon^2$ gives
\begin{equation}\label{f_2}\displaystyle
f_2=\sigma_1\sum_{i\leq i,j\leq 2}e^{\theta_i+\bar{\theta}_j}\alpha_{ij}+\sigma_2\sum_{i\leq i,j\leq 2}e^{\eta_i+\bar{\eta}_j}\beta_{ij},
\end{equation}
where
\begin{equation}\displaystyle
\alpha_{ij}=\frac{1}{2(k_i+\bar{k}_j)(l_i+\bar{l}_j)},\quad \beta_{ij}=\frac{1}{2(r_i+\bar{r}_j)(s_i+\bar{s}_j)}, \quad i\leq i,j\leq 2.
\end{equation}
From the coefficients of $\epsilon^3$ we obtain the functions $g_3$ and $h_3$ as
\begin{align}
&g_3=A_1e^{\theta_1+\eta_2+\bar{\eta}_1}+A_2e^{\theta_2+\eta_1+\bar{\eta}_1}+A_3e^{\theta_1+\eta_2+\bar{\eta}_2}+A_4e^{\theta_2+\eta_1+\bar{\eta}_2}+
A_5e^{\theta_1+\theta_2+\bar{\theta}_1}+A_6e^{\theta_1+\theta_2+\bar{\theta}_2},\nonumber\\
&h_3=B_1e^{\theta_2+\eta_1+\bar{\theta}_1}+B_2e^{\theta_1+\eta_2+\bar{\theta}_1}+B_3e^{\theta_2+\eta_1+\bar{\theta}_2}+B_4e^{\theta_1+\eta_2+\bar{\theta}_2}+
B_5e^{\eta_1+\eta_2+\bar{\eta}_1}+B_6e^{\eta_1+\eta_2+\bar{\eta}_2},\label{g_3h_3}
\end{align}
where{\small
\begin{align*}
&A_1=\sigma_2\beta_{21}\frac{(k_1-k_2)}{(k_1+\bar{k}_1)},\, A_2=-\sigma_2\beta_{11}\frac{(k_1-k_2)}{(k_2+\bar{k}_1)},\,
A_3=\sigma_2\beta_{22}\frac{(k_1-k_2)}{(k_1+\bar{k}_2)},\, A_4=-\sigma_2\beta_{12}\frac{(k_1-k_2)}{(k_2+\bar{k}_2)},\\\\
&A_5=-\sigma_1(k_1-k_2)\frac{[\alpha_{11}(k_1+\bar{k}_1)-\alpha_{21}(k_2+\bar{k}_1)]}{(k_1+\bar{k}_1)(k_2+\bar{k}_1)},\,
A_6=-\sigma_1(k_1-k_2)\frac{[\alpha_{12}(k_1+\bar{k}_2)-\alpha_{22}(k_2+\bar{k}_2)]}{(k_1+\bar{k}_2)(k_2+\bar{k}_2)},
\end{align*}}
and{\small
\begin{align*}
&B_1=\sigma_1\alpha_{21}\frac{(k_1-k_2)}{(k_1+\bar{k}_1)},\, B_2=-\sigma_1\alpha_{11}\frac{(k_1-k_2)}{(k_2+\bar{k}_1)},\,
B_3=\sigma_1\alpha_{22}\frac{(k_1-k_2)}{(k_1+\bar{k}_2)},\, B_4=-\sigma_1\alpha_{12}\frac{(k_1-k_2)}{(k_2+\bar{k}_2)},\\\\
&B_5=-\sigma_2(k_1-k_2)\frac{[\beta_{11}(k_1+\bar{k}_1)-\beta_{21}(k_2+\bar{k}_1)]}{(k_1+\bar{k}_1)(k_2+\bar{k}_1)},\,
B_6=-\sigma_2(k_1-k_2)\frac{[\beta_{12}(k_1+\bar{k}_2)-\beta_{22}(k_2+\bar{k}_2)]}{(k_1+\bar{k}_2)(k_2+\bar{k}_2)},
\end{align*}}
with the following conditions satisfied:
\begin{equation}\label{constraintstwo}
a=-\bar{a},\quad r_j=k_j,\quad j=1, 2.
\end{equation}
The coefficient of $\epsilon^4$ yields the function $f_4$ as{\small
\begin{align}\label{f_4}
f_4=&C_1e^{\theta_1+\bar{\theta}_1+\eta_2+\bar{\eta}_2}+C_2e^{\theta_2+\bar{\theta}_2+\eta_1+\bar{\eta}_1}+C_3e^{\eta_1+\bar{\eta}_1+\eta_2+\bar{\eta}_2}\nonumber\\
&+C_4e^{\theta_1+\bar{\theta}_1+\theta_2+\bar{\theta}_2}+C_5e^{\theta_1+\bar{\theta}_2+\eta_2+\bar{\eta}_1}+C_6e^{\theta_2+\bar{\theta}_1+\eta_1+\bar{\eta}_2},
\end{align}}
where{\small
\begin{align*}
&C_1=\frac{\sigma_1\sigma_2\kappa}{(s_2+\bar{s}_2)(l_1+\bar{l}_1)},\, C_2=\frac{\sigma_1\sigma_2\kappa}{(s_1+\bar{s}_1)(l_2+\bar{l}_2)},\,
C_3=\frac{\sigma_2^2\kappa(s_1-s_2)(\bar{s}_1-\bar{s}_2)}{(s_1+\bar{s}_1)(s_1+\bar{s}_2)(s_2+\bar{s}_1)(s_2+\bar{s}_2)},\nonumber\\\\
&C_4=\frac{\sigma_1^2\kappa(l_1-l_2)(\bar{l}_1-\bar{l}_2)}{(l_1+\bar{l}_1)(l_1+\bar{l}_2)(l_2+\bar{l}_1)(l_2+\bar{l}_2)},\,
C_5=-\frac{\sigma_1\sigma_2\kappa}{(s_2+\bar{s}_1)(l_1+\bar{l}_2)},\,C_6=-\frac{\sigma_1\sigma_2\kappa}{(s_1+\bar{s}_2)(l_2+\bar{l}_1)},
\end{align*}}
for
\begin{equation}
\kappa=\frac{(k_1-k_2)(\bar{k}_1-\bar{k}_2)}{4(k_1+\bar{k}_1)(k_1+\bar{k}_2)(k_2+\bar{k}_1)(k_2+\bar{k}_2)}.
\end{equation}
The coefficients of $\epsilon^j$, $j=5, 6, 7, 8$ vanish directly. Take also $\epsilon=1$. Since two-soliton solution of the $3$-component Maccari system (\ref{N=2-a})-(\ref{N=2-c})
has a long expression, we will not give it here explicitly. It can be expressed as $(u(x,y,t),v(x,y,t),p(x,y,t))$ where
\begin{equation}\label{twosoliton}
u=\frac{g_1+g_3}{1+f_2+f_4},\quad v=\frac{h_1+h_3}{1+f_2+f_4},\quad p=2(\ln(1+f_2+f_4))_{xx}.
\end{equation}
Here $g_1, h_1$ are given in (\ref{g_1h_1}), $g_3, h_3$ in (\ref{g_3h_3}), $f_2$ in (\ref{f_2}), and $f_4$ in (\ref{f_4}) with
the dispersion relations (\ref{dispersiontwosoliton}) and the constraints (\ref{constraintstwo}) satisfied.

\section{Local reductions of $3$-component Maccari system}

\noindent \textbf{(a)}\, $v(x,y,t)=\rho u(x,y,t)$, $\rho$ is a real constant.

The system (\ref{N=2-a})-(\ref{N=2-c}) consistently reduces to the local system
\begin{align}
&au_t+u_{xx}+pu=0,\label{N=2local1-a}\\
&p_y=(\sigma_1+\sigma_2\rho^2)(u\bar{u})_x.\label{N=2local1-b}
\end{align}
Here there is no restriction on the constant $a$.

\noindent \textbf{(b)}\, $v(x,y,t)=\rho \bar{u}(x,y,t)$, $\rho$ is a real constant.

In this case the system (\ref{N=2-a})-(\ref{N=2-c}) consistently reduces to the same local system
(\ref{N=2local1-a}) and (\ref{N=2local1-b}) with the restriction $a=\bar{a}$.

\section{Nonlocal reductions of $3$-component Maccari system}

\noindent \textbf{(a)}\, $v(x,y,t)=\rho u(\varepsilon_1x,\varepsilon_2y,\varepsilon_3t)=\rho u^{\varepsilon}$, $\rho$ is a real constant, and $\varepsilon_j=\pm 1$, $j=1,2,3$.

We apply this reduction to $3$-component Maccari system (\ref{N=2-a})-(\ref{N=2-c}). We get
\begin{align}
au_t+u_{xx}+uD_y^{-1}[\sigma_1u\bar{u}+\sigma_2\rho^2u^{\varepsilon}\bar{u}^{\varepsilon}]_x=0,\label{rneq1}\\
a\rho u_t^{\varepsilon}+\rho u_{xx}^{\varepsilon}+\rho u^{\varepsilon} D_y^{-1}[\sigma_1u\bar{u}+\sigma_2\rho^2u^{\varepsilon}\bar{u}^{\varepsilon}]_x=0.\label{rneq2}
\end{align}
Here $D_y^{-1}$ is the standard anti-derivative with respect to $y$. Let $x\rightarrow \varepsilon_1 x$, $y\rightarrow \varepsilon_2 y$, and $t\rightarrow \varepsilon_3 t$ in (\ref{rneq2}). We have
\begin{equation}
a\varepsilon_3u_t+u_{xx}+\varepsilon_1\varepsilon_2uD_y^{-1}[\sigma_1u^{\varepsilon}\bar{u}^{\varepsilon}+\sigma_2\rho^2u\bar{u}]_x=0.
\end{equation}
For consistency, this equation must be same with (\ref{rneq1}) provided that
\begin{equation}
 \varepsilon_3=1, \quad  \varepsilon_1\varepsilon_2\sigma_1=\sigma_2\rho^2.
\end{equation}
Since $\sigma_j=\pm 1$, $j=1,2$, we have $\rho=\pm 1$.
Then the system (\ref{N=2-a})-(\ref{N=2-c}) reduces to S-reversal nonlocal systems consistently,
\begin{align}
&au_t(x,y,t)+u_{xx}(x,y,t)+p(x,y,t)u(x,y,t)=0,\label{N=2nonlocal1-a}\\
&p_y(x,y,t)=(\sigma_1u(x,y,t)\bar{u}(x,y,t)+\sigma_2u(\varepsilon_1x,\varepsilon_2y,t)\bar{u}(\varepsilon_1x,\varepsilon_2y,t))_x.\label{N=2nonlocal1-b}
\end{align}

Explicitly, we have three different S-reversal nonlocal systems:\\

\noindent \textbf{(i)}\, $(\varepsilon_1,\varepsilon_2,\varepsilon_3)=(1,-1,1), \sigma_2=-\sigma_1, \sigma_1=\pm 1$.
\begin{align}
&au_t(x,y,t)+u_{xx}(x,y,t)+p(x,y,t)u(x,y,t)=0,\label{realnon1-a}\\
&p_y(x,y,t)=(\sigma_1u(x,y,t)\bar{u}(x,y,t)+\sigma_2u(x,-y,t)\bar{u}(x,-y,t))_x.\label{realnon1-b}
\end{align}

\noindent \textbf{(ii)}\, $(\varepsilon_1,\varepsilon_2,\varepsilon_3)=(-1,1,1), \sigma_2=-\sigma_1, \sigma_1=\pm 1$.
\begin{align}
&au_t(x,y,t)+u_{xx}(x,y,t)+p(x,y,t)u(x,y,t)=0,\label{realnon2-a}\\
&p_y(x,y,t)=(\sigma_1u(x,y,t)\bar{u}(x,y,t)+\sigma_2u(-x,y,t)\bar{u}(-x,y,t))_x.\label{realnon2-b}
\end{align}

\noindent \textbf{(iii)}\, $(\varepsilon_1,\varepsilon_2,\varepsilon_3)=(-1,-1,1), \sigma_2=\sigma_1, \sigma_1=\pm 1$.
\begin{align}
&au_t(x,y,t)+u_{xx}(x,y,t)+p(x,y,t)u(x,y,t)=0,\label{realnon3-a}\\
&p_y(x,y,t)=(\sigma_1u(x,y,t)\bar{u}(x,y,t)+\sigma_2u(-x,-y,t)\bar{u}(-x,-y,t))_x.\label{realnon3-b}
\end{align}

\noindent \textbf{(b)}\, $v(x,y,t)=\rho \bar{u}(\varepsilon_1x,\varepsilon_2y,\varepsilon_3t)$, $\rho$ is a real constant, and $\varepsilon_k=\pm 1$, $k=1,2,3$.

\noindent When we apply this nonlocal reduction to the system (\ref{N=2-a})-(\ref{N=2-c}) and making the similar analysis as in part (a), we get the following constraints for consistency:
\begin{equation}
 a=\bar{a}\varepsilon_3,\quad  \varepsilon_1\varepsilon_2\sigma_1=\sigma_2\rho^2,
\end{equation}
yielding $\rho=\pm 1$. The system (\ref{N=2-a})-(\ref{N=2-c}) reduces to the following nonlocal systems:
\begin{align}
&au_t(x,y,t)+u_{xx}(x,y,t)+p(x,y,t)u(x,y,t)=0,\label{N=2nonlocal2-a}\\
&p_y(x,y,t)=(\sigma_1u(x,y,t)\bar{u}(x,y,t)+\sigma_2u(\varepsilon_1x,\varepsilon_2y,\varepsilon_3t)\bar{u}(\varepsilon_1x,\varepsilon_2y,\varepsilon_3t))_x.\label{N=2nonlocal2-b}
\end{align}
\noindent Explicitly, here we have seven different nonlocal T-, S-, and ST-reversal $2$-component Maccari systems corresponding to

\noindent \textbf{(i)}\, $(\varepsilon_1,\varepsilon_2,\varepsilon_3)=(1,1,-1), \sigma_2=\sigma_1$, $a=-\bar{a}$,

\noindent \textbf{(ii)}\, $(\varepsilon_1,\varepsilon_2,\varepsilon_3)=(1,-1,-1), \sigma_2=-\sigma_1$, $a=-\bar{a}$,

\noindent \textbf{(iii)}\, $(\varepsilon_1,\varepsilon_2,\varepsilon_3)=(-1,-1,-1), \sigma_2=\sigma_1$, $a=-\bar{a}$,

\noindent \textbf{(iv)}\, $(\varepsilon_1,\varepsilon_2,\varepsilon_3)=(-1,1,-1), \sigma_2=-\sigma_1$, $a=-\bar{a}$,

\noindent \textbf{(v)}\, $(\varepsilon_1,\varepsilon_2,\varepsilon_3)=(1,-1,1), \sigma_2=-\sigma_1$, $a=\bar{a}$,

\noindent \textbf{(vi)}\, $(\varepsilon_1,\varepsilon_2,\varepsilon_3)=(-1,-1,1), \sigma_2=\sigma_1$, $a=\bar{a}$,

\noindent \textbf{(vii)}\, $(\varepsilon_1,\varepsilon_2,\varepsilon_3)=(-1,1,1), \sigma_2=-\sigma_1$, $a=\bar{a}$.\\

\noindent Note that the systems obtained for the cases (v)-(vii) have already been given in the part (a) with no restriction on the constant $a$.

\noindent In \cite{Mihalache}, the $(2+1)$-dimensional nonlocal Fokas system given by
\begin{align}
&iu_t(x,y,t)+u_{xx}(x,y,t)+u(x,y,t)p(x,y,t)=0,\nonumber\\
&p_y=(u(x,y,t)\bar{u}(-x,-y,t))_x,
\end{align}
was studied and semi-rational solutions of this system were obtained. The above system is similar but different than the system (\ref{N=2nonlocal2-a}) and (\ref{N=2nonlocal2-b}) corresponding to (vi).

\section{Soliton solutions of the local and nonlocal reduced systems}

In Sections 3 and 4 we have obtained consistent local and nonlocal reductions of $3$-component Maccari system (\ref{N=2-a})-(\ref{N=2-c}) and
the constraints to be satisfied. By using the reduction formulas with one- and two-soliton solutions of (\ref{N=2-a})-(\ref{N=2-c}) and following Type 1 or Type 2 approaches \cite{GurPek1}, \cite{GurPek3}, \cite{GurPek2} we can obtain one- and two-soliton solutions of the local and nonlocal reduced $2$-component Maccari systems.

\subsection{One-soliton solutions of the local reduced systems}

Both of the local reductions; $v(x,y,t)=\rho u(x,y,t)$ and $v(x,y,t)=\rho \bar{u}(x,y,t)$, yield the same reduced local system (\ref{N=2local1-a}) and (\ref{N=2local1-b}). The latter
one puts a condition on the constant $a$ as $a=\bar{a}$. Note that we have given one-soliton solution of $3$-component Maccari system (\ref{N=2-a})-(\ref{N=2-c})
under the constraints (\ref{onesolcond}). Therefore we only consider the first local reduction $v(x,y,t)=\rho u(x,y,t)$ giving
\begin{equation}
\frac{h}{f}=\rho\frac{g}{f}.
\end{equation}
When we use Type 1 that is the approach based on equating numerators and denominators separately, we get the following conditions:
\begin{equation}
l_2=l_1,\quad e^{\delta_2}=\rho e^{\delta_1}.
\end{equation}
Hence one-soliton solution of the local reduced system (\ref{N=2local1-a}) and (\ref{N=2local1-b}) is given by the pair $(u(x,y,t),p(x,y,t))$ where
\begin{equation}\label{localsolorig}
u(x,y,t)=\frac{e^{k_1x+\omega_1t+l_1y+\delta_1}}{(1+\frac{(\sigma_1+\rho^2\sigma_2)}{2(k_1+\bar{k}_1)(l_1+\bar{l}_1)}e^{(k_1+\bar{k}_1)x
+(\omega_1+\bar{\omega}_1)t+(l_1+\bar{l}_1)y+\delta_1+\bar{\delta}_1})}
\end{equation}
and
\begin{equation}\label{localp(x,y,t)}
p(x,y,t)=\frac{(k_1+\bar{k}_1)(\sigma_1+\rho^2\sigma_2)e^{(k_1+\bar{k}_1)x
+(\omega_1+\bar{\omega}_1)t+(l_1+\bar{l}_1)y+\delta_1+\bar{\delta}_1}   }{(l_1+\bar{l}_1)(1+\frac{(\sigma_1+\rho^2\sigma_2)}{2(k_1+\bar{k}_1)(l_1+\bar{l}_1)}e^{(k_1+\bar{k}_1)x
+(\omega_1+\bar{\omega}_1)t+(l_1+\bar{l}_1)y+\delta_1+\bar{\delta}_1})^2}.
\end{equation}
\smallskip
Here $\omega_1=-\frac{k_1^2}{a}$, $\sigma_j=\pm 1$ for $j=1, 2$, and $a=-\bar{a}$. Let $k_1=\alpha_1+i\beta_1$, $\omega_1=\alpha_2+i\beta_2$, $a=i\beta_3$, $l_1=\alpha_4+i\beta_4$, and $e^{\delta_1}=\alpha_5+i\beta_5$ where $\alpha_1, \alpha_2, \alpha_4,\alpha_5 \in \mathbb{R}$, $\beta_j\in \mathbb{R}$ for $1\leq j \leq 5$. Indeed, $\alpha_2=-\frac{2\alpha_1\beta_1}{\beta_3}$ and $\beta_2=\frac{(\alpha_1^2-\beta_1^2)}{\beta_3}$.
Then from the solution (\ref{localsolorig}) we have
\begin{equation}
|u(x,y,t)|^2=\frac{2\alpha_1\alpha_4}{(\sigma_1+\rho^2\sigma_2)}\mathrm{sech}^2(\alpha_1x+\alpha_2t+\alpha_4y+\phi)
\end{equation}
and from (\ref{localp(x,y,t)}) we get
\begin{equation}
p(x,y,t)=2\alpha_1^2\mathrm{sech}^2(\alpha_1x+\alpha_2t+\alpha_4y+\phi)
\end{equation}
\smallskip
with $\phi=\frac{1}{2}\ln|\frac{(\alpha_5^2+\beta_5^2)(\sigma_1+\rho^2\sigma_2)}{8\alpha_1\alpha_4}|$. These solutions are nonsingular and bounded if $\frac{(\sigma_1+\rho^2\sigma_2)}{\alpha_1\alpha_4}>0$. They are bell-shaped soliton solutions.

\noindent \textbf{Example 1.} Take the parameters of one-soliton solution of the local system (\ref{N=2local1-a}) and (\ref{N=2local1-b}) as $(k_1,l_1,a,\rho)=(1+\frac{i}{2},\frac{1}{3}, 2i,4)$ with $\sigma_1=\sigma_2=e^{\delta_1}=1$. Then the pair $(|u(x,y,t)|^2,p(x,y,t))$ becomes
\begin{equation}\label{ex1}
|u(x,y,t)|^2=\frac{2}{51}\mathrm{sech}^2(x+\frac{1}{3}y-\frac{1}{2}t+\delta),\quad p(x,y,t)=2\mathrm{sech}^2(x+\frac{1}{3}y-\frac{1}{2}t+\frac{1}{2}+\delta),
\end{equation}
where $\delta=\frac{1}{2}\ln(\frac{51}{8})$. The graphs of the above solutions at $t=0$ are given in Figure 1.
\begin{center}
\begin{figure}[h!]
\centering
\subfloat[]{\includegraphics[width=0.35\textwidth]{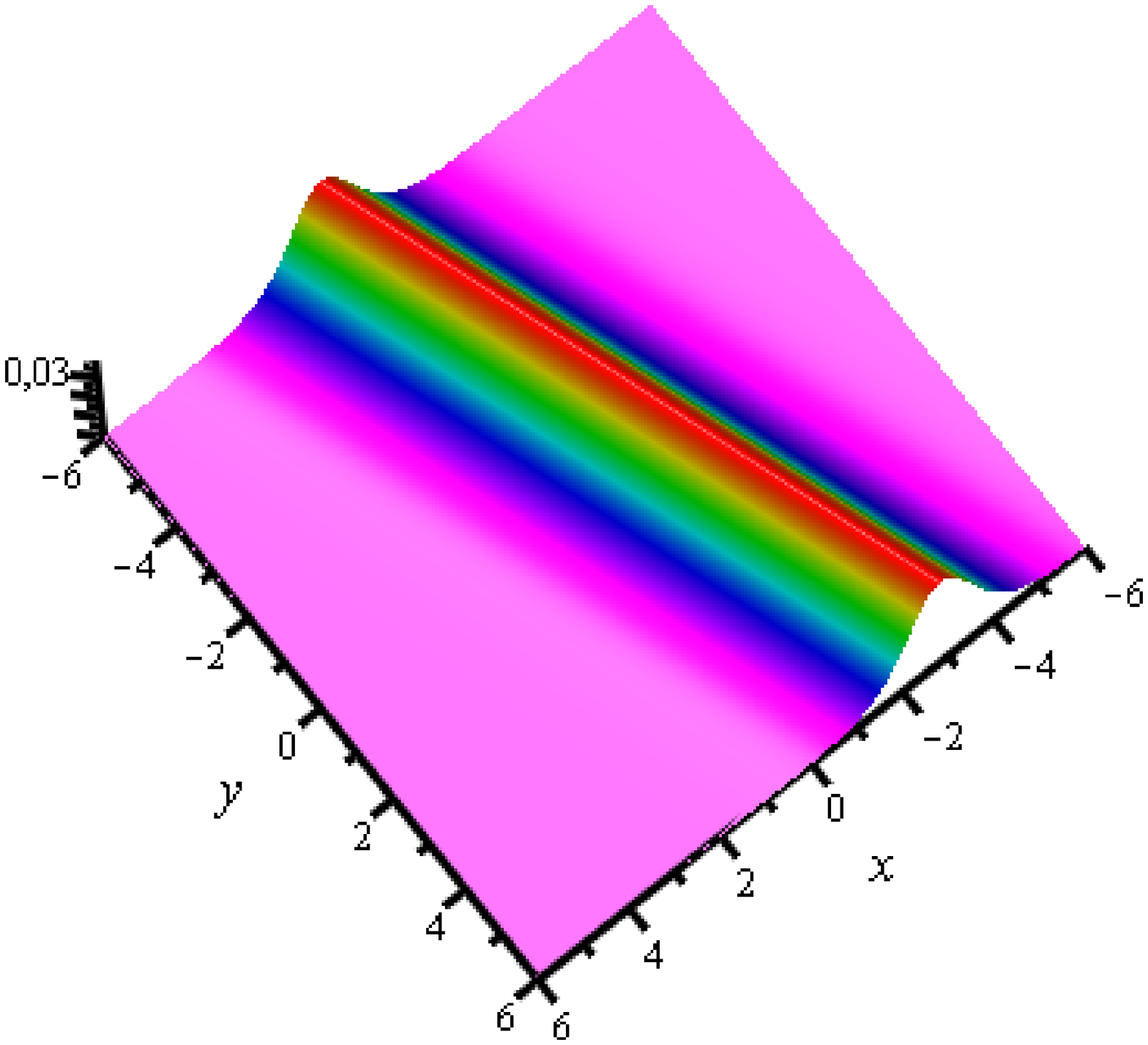}}\hspace{2cm}
\subfloat[] {\includegraphics[width=0.35\textwidth]{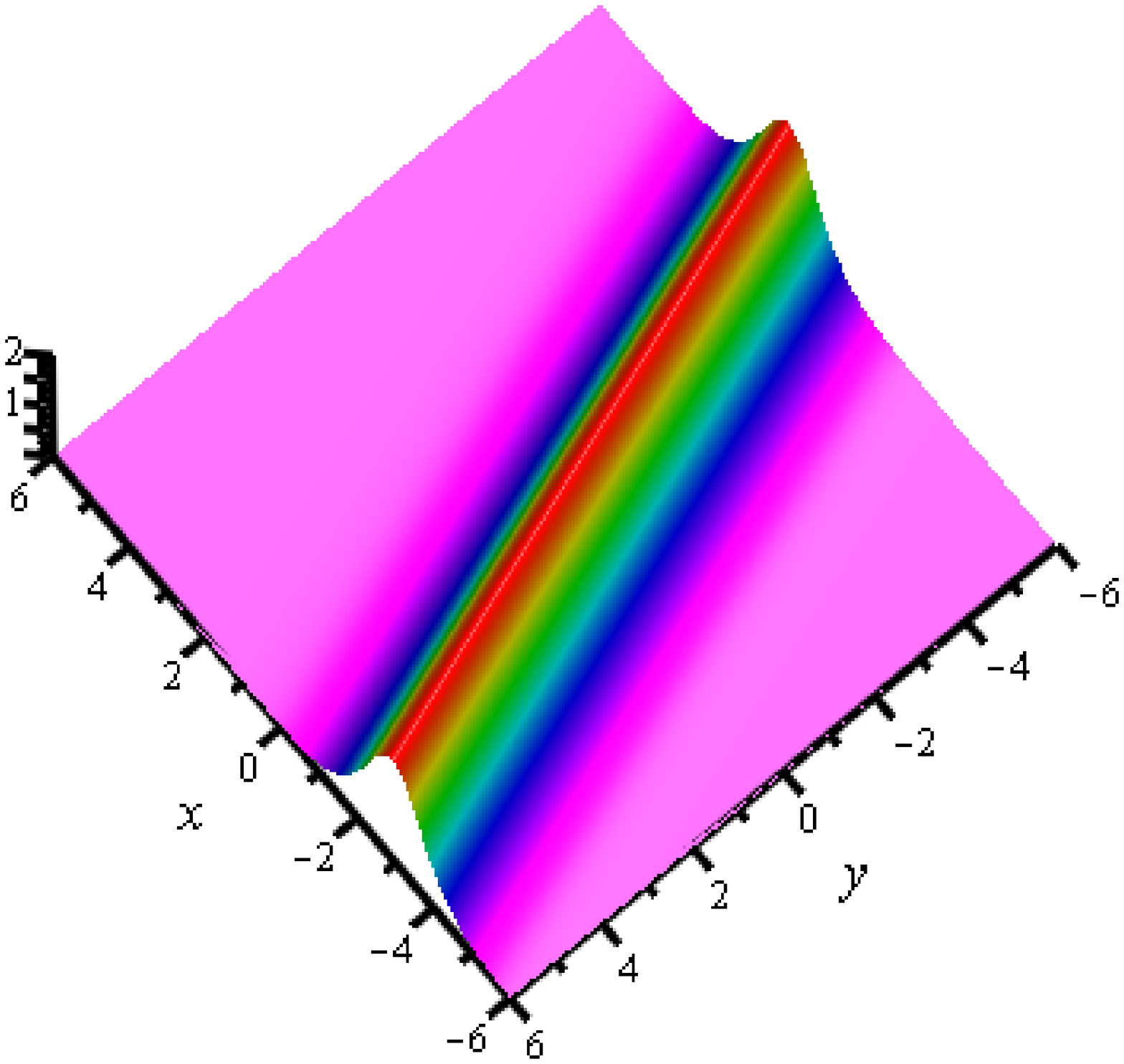}}
\caption{One-soliton solutions of the local system (\ref{N=2local1-a}) and (\ref{N=2local1-b}) at $t=0$ for the parameters $(k_1,l_1,a,\rho)=(1+\frac{i}{2},\frac{1}{3}, 2i,4)$ with $\sigma_1=\sigma_2=e^{\delta_1}=1$ (a) $|u(x,y,t)|^2$, (b) $p(x,y,t)$.}
\end{figure}
\end{center}
\squeezeup

Figures 2(a) and 2(b) depict the movements of the solitons from $|u(x,y,t)|^2$ and $p(x,y,t)$ for different times in different positions at the $y$-axis, respectively.
\newpage
\begin{center}
\begin{figure}[h!]
\centering
\subfloat[]{\includegraphics[width=0.29\textwidth]{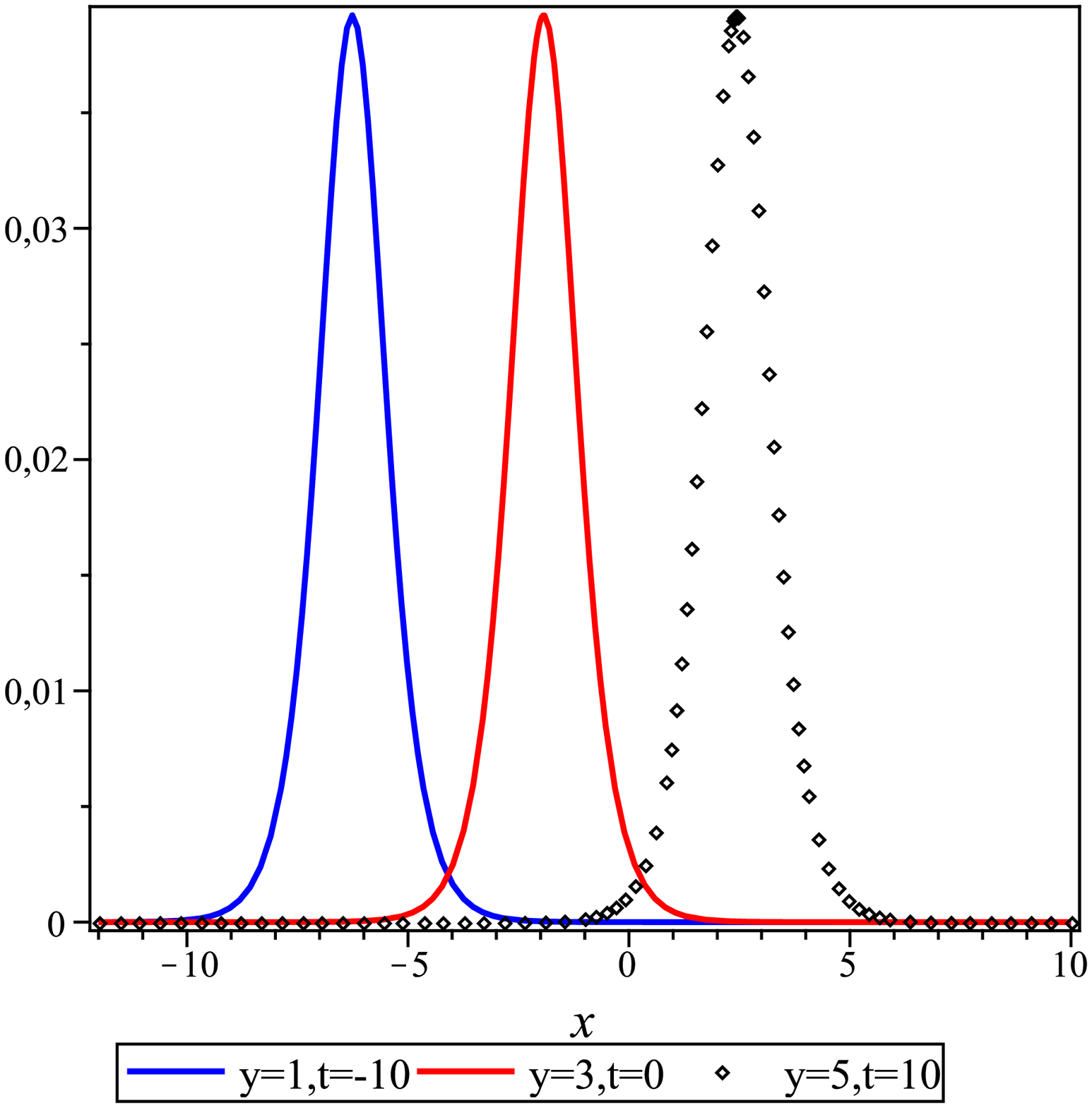}}\hspace{2cm}
\subfloat[] {\includegraphics[width=0.29\textwidth]{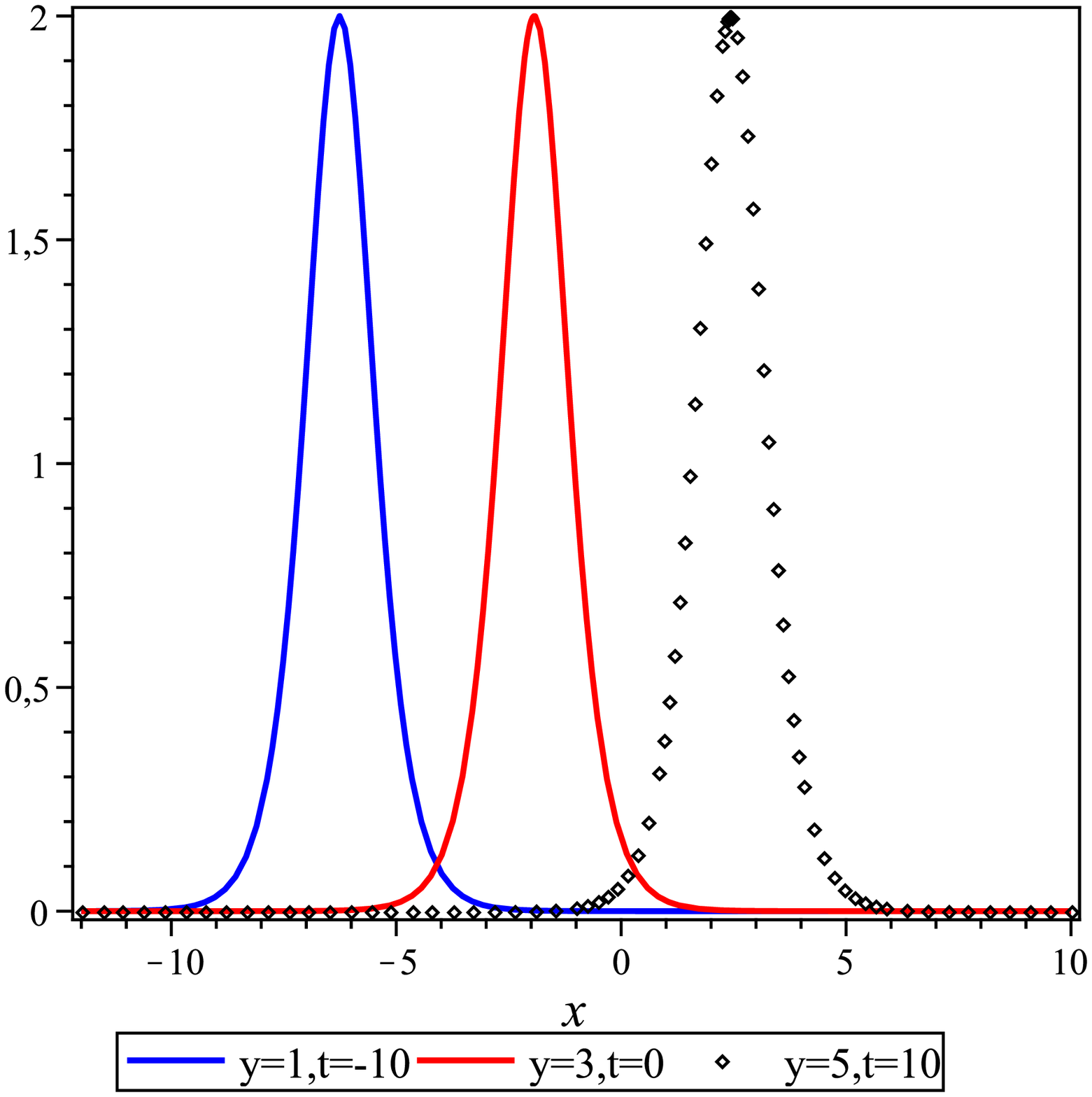}}
\caption{The movement of the waves (\ref{ex1}) as time increases.}
\end{figure}
\end{center}
\squeezeup
\subsection{One-soliton solutions of the nonlocal reduced systems}

In this section we present one-soliton solutions of the nonlocal systems obtained from $3$-component Maccari system (\ref{N=2-a})-(\ref{N=2-c}) by two types of nonlocal reductions.

\noindent \textbf{(a)}\, $v(x,y,t)=\rho u(\varepsilon_1x,\varepsilon_2y,\varepsilon_3t)$, $\rho=\pm 1$, and $\varepsilon_j=\pm 1$, $j=1,2,3$.

By this nonlocal reduction we obtain three types of nonlocal systems corresponding to $(\varepsilon_1,\varepsilon_2,\varepsilon_3)=(1,-1,1)$ with $\sigma_2=-\sigma_1$; $(\varepsilon_1,\varepsilon_2,\varepsilon_3)=(-1,1,1)$ with $\sigma_2=-\sigma_1$; and $(\varepsilon_1,\varepsilon_2,\varepsilon_3)=(-1,-1,1)$ with $\sigma_2=\sigma_1$. We analyze these cases separately.

\noindent \textbf{(a).(i)} $(\varepsilon_1,\varepsilon_2,\varepsilon_3)=(1,-1,1), \sigma_2=-\sigma_1$, $\sigma_1=\pm 1$.

\noindent When we use Type 1, the reduction formula yields the conditions,
\begin{equation}
l_2=-l_1,\quad e^{\delta_2}=\rho e^{\delta_1}.
\end{equation}
Therefore one-soliton solution of the nonlocal system (\ref{realnon1-a}) and (\ref{realnon1-b}) is the pair $(u(x,y,t), p(x,y,t))$ where
\begin{equation}\label{nonlocalsolreali}
u(x,y,t)=\frac{e^{k_1x+\omega_1 t+l_1y+\delta_1}}{1+\sigma_1\frac{e^{(k_1+\bar{k}_1)x+(\omega_1+\bar{\omega}_1)t+\delta_1+\bar{\delta}_1}}{(k_1+\bar{k}_1)(l_1+\bar{l}_1)}\cosh((l_1+\bar{l}_1)y)}
\end{equation}
and $p(x,y,t)$ is
\begin{equation}
p(x,y,t)=\frac{ \frac{2\sigma_1(k_1+\bar{k}_1)}{(l_1+\bar{l}_1)}e^{(k_1+\bar{k}_1)x+(\omega_1+\bar{\omega}_1)t+\delta_1+\bar{\delta}_1}\cosh((l_1+\bar{l}_1)y)     }{[1+\frac{\sigma_1}{(k_1+\bar{k}_1)(l_1+\bar{l}_1)}e^{(k_1+\bar{k}_1)x+(\omega_1+\bar{\omega}_1)t+\delta_1+\bar{\delta}_1}\cosh((l_1+\bar{l}_1)y)]^2}.
\end{equation}
 Let $k_1=\alpha_1+i\beta_1$, $\omega_1=\alpha_2+i\beta_2$, $a=i\beta_3$, $l_1=\alpha_4+i\beta_4$, and $e^{\delta_1}=\alpha_5+i\beta_5$ where $\alpha_1, \alpha_2, \alpha_4,\alpha_5 \in \mathbb{R}$, $\beta_j\in \mathbb{R}$ for $1 \leq j\leq 5$. Here $\alpha_2=-\frac{2\alpha_1\beta_1}{\beta_3}$ and $\beta_2=\frac{(\alpha_1^2-\beta_1^2)}{\beta_3}$. Then solutions become
\begin{equation}\label{nonlocalsolrealimodulus}
|u(x,y,t)|^2=\frac{e^{2\alpha_1x+2\alpha_2t+2\alpha_4y}(\alpha_5^2+\beta_5^2)}{(1+\frac{\sigma_1(\alpha_5^2+\beta_5^2)}{4\alpha_1\alpha_4}e^{2\alpha_1x+2\alpha_2t}
\cosh(2\alpha_4y))^2},
\end{equation}
and
\begin{equation}
p(x,y,t)=\frac{2\alpha_1(\alpha_5^2+\beta_5^2)\sigma_1e^{2\alpha_1x+2\alpha_2t}\cosh(2\alpha_4y)}{\alpha_4[1+\frac{\sigma_1}{4\alpha_1\alpha_4}(\alpha_5^2+\beta_5^2)
 e^{2\alpha_1x+2\alpha_2t}\cosh(2\alpha_4y)]^2}.
\end{equation}
The above solutions are nonsingular and bounded if $\frac{\sigma_1}{\alpha_1\alpha_4}>0$. Consider the following particular example.

\noindent \textbf{Example 2.} Choose the parameters of the one-soliton solution (\ref{nonlocalsolreali}) as $(k_1,l_1,a,\sigma_1,e^{\delta_1},\rho)=(1+\frac{i}{4},\frac{1}{2},i,1,1,1)$. Then the solutions of the nonlocal system (\ref{realnon1-a}) and (\ref{realnon1-b}) become
\begin{equation}\label{example2up}
|u(x,y,t)|^2=\frac{e^{2x-t+y}}{(1+\frac{1}{2}e^{2x-t}\cosh(y))^2},\quad
p(x,y,t)=\frac{4e^{2x-t}\cosh(y)}{(1+\frac{1}{2}e^{2x-t}\cosh(y))^2}.
\end{equation}
Both of the above solutions are nonsingular and bounded. The graphs of (\ref{example2up}) at $t=0$ are given in Figure 3.
\begin{center}
\begin{figure}[h!]
\centering
\subfloat[]{\includegraphics[width=0.35\textwidth]{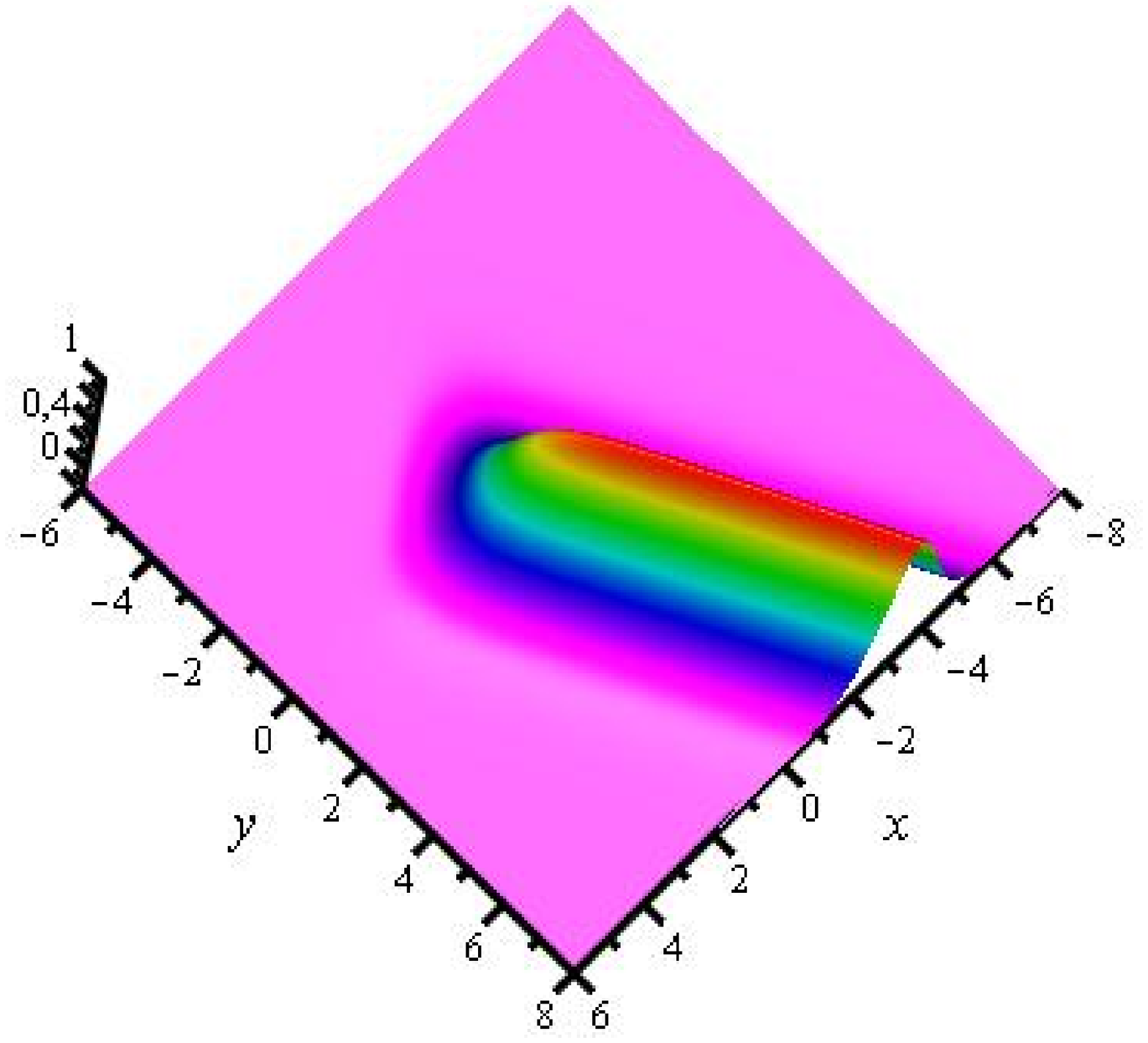}}\hspace{2cm}
\subfloat[] {\includegraphics[width=0.35\textwidth]{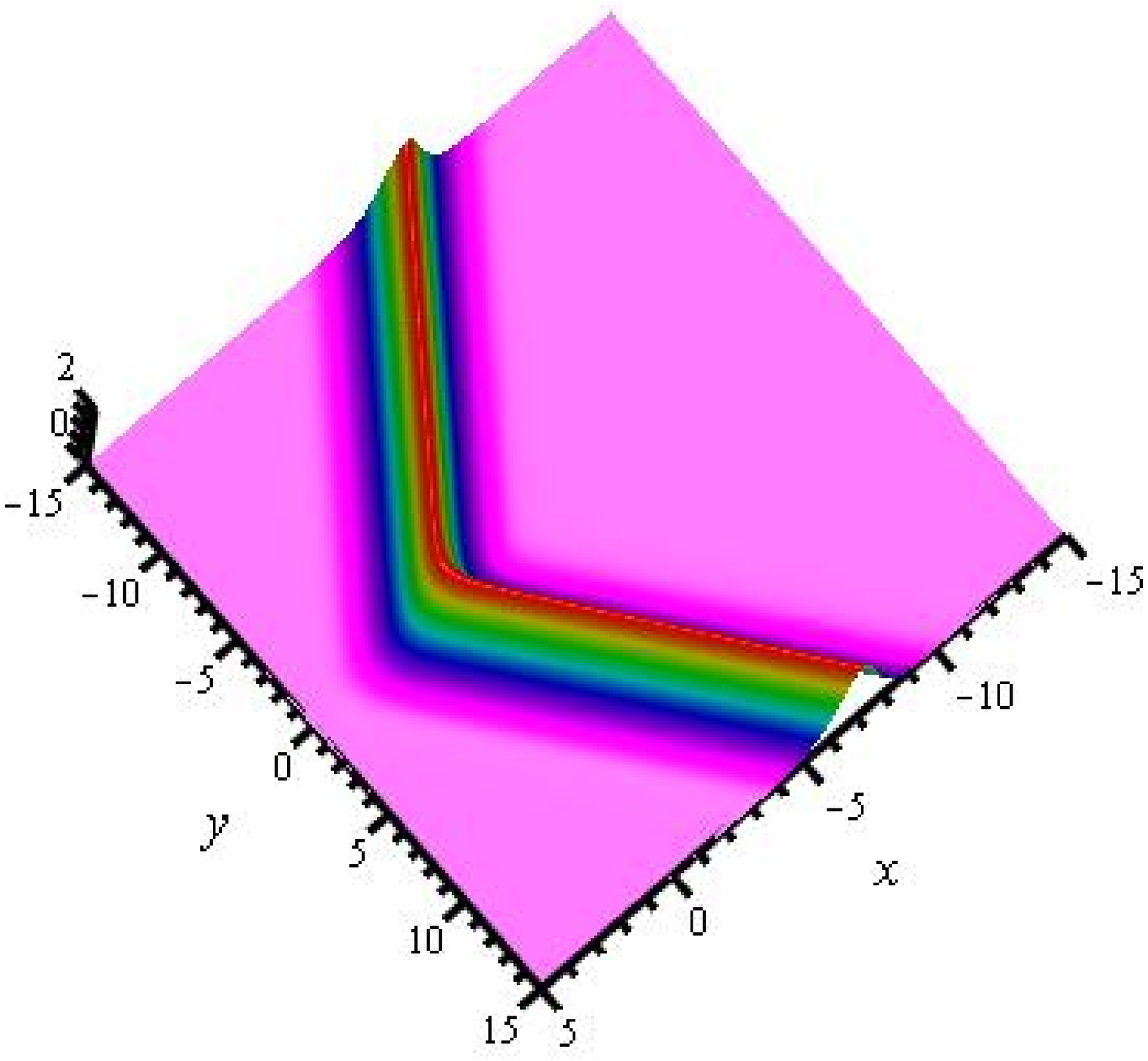}}
\caption{One-soliton solutions of the nonlocal system (\ref{realnon1-a}) and (\ref{realnon1-b}) at $t=0$ for the parameters $k_1=1+\frac{i}{4}, l_1=\frac{1}{2}, a=i, \sigma_1=e^{\delta_1}=\rho=1$. (a) Solitoff solution $|u(x,y,t)|^2$, (b) V-type solitary wave solution $p(x,y,t)$.}
\end{figure}
\end{center}
\squeezeup

Figures 4(a) and 4(b) show the movements of the waves from $|u(x,y,t)|^2$ and $p(x,y,t)$ for different times in different positions at the $y$-axis, respectively.
\begin{center}
\begin{figure}[h!]
\centering
\subfloat[]{\includegraphics[width=0.29\textwidth]{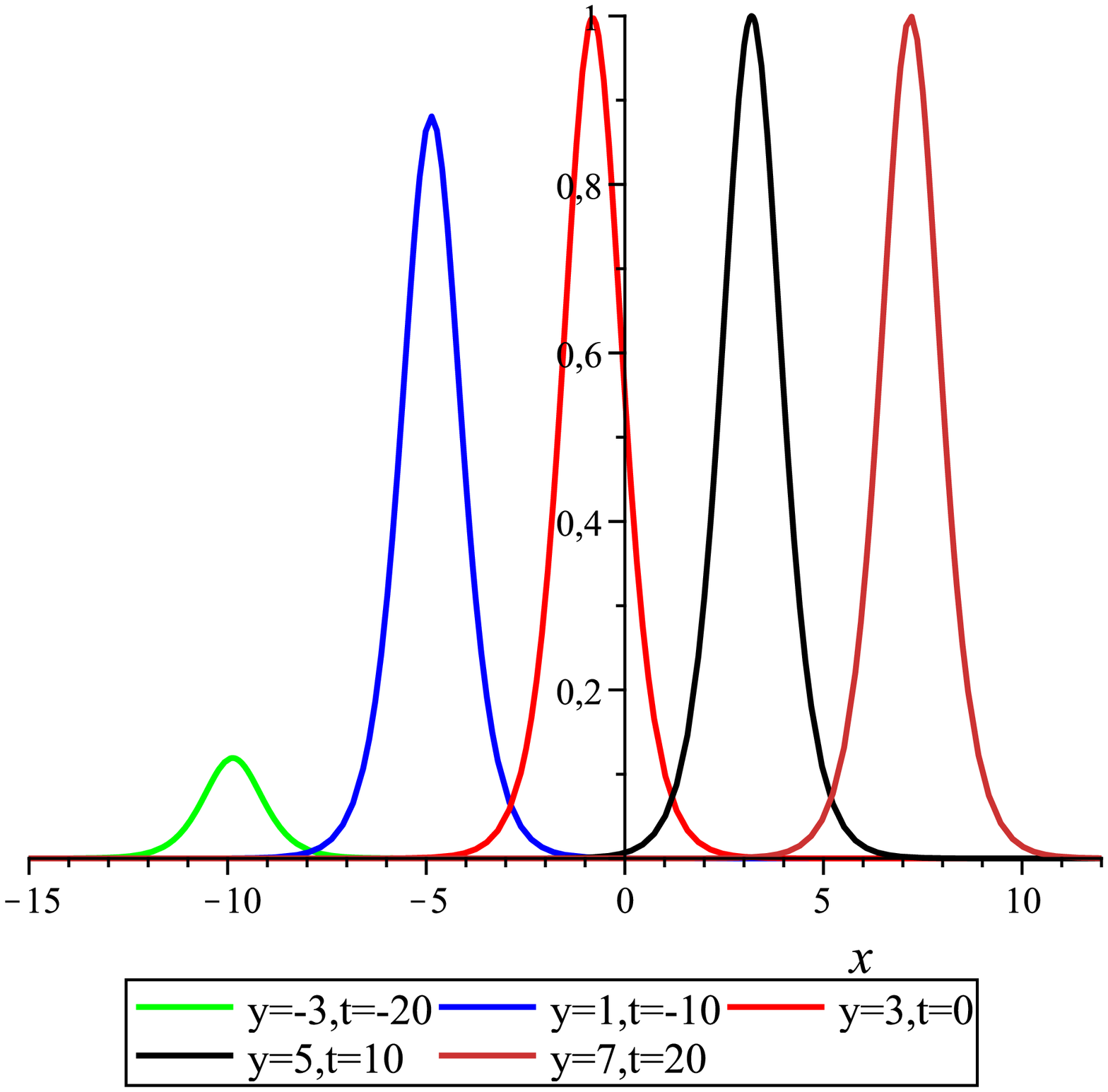}}\hspace{2cm}
\subfloat[] {\includegraphics[width=0.29\textwidth]{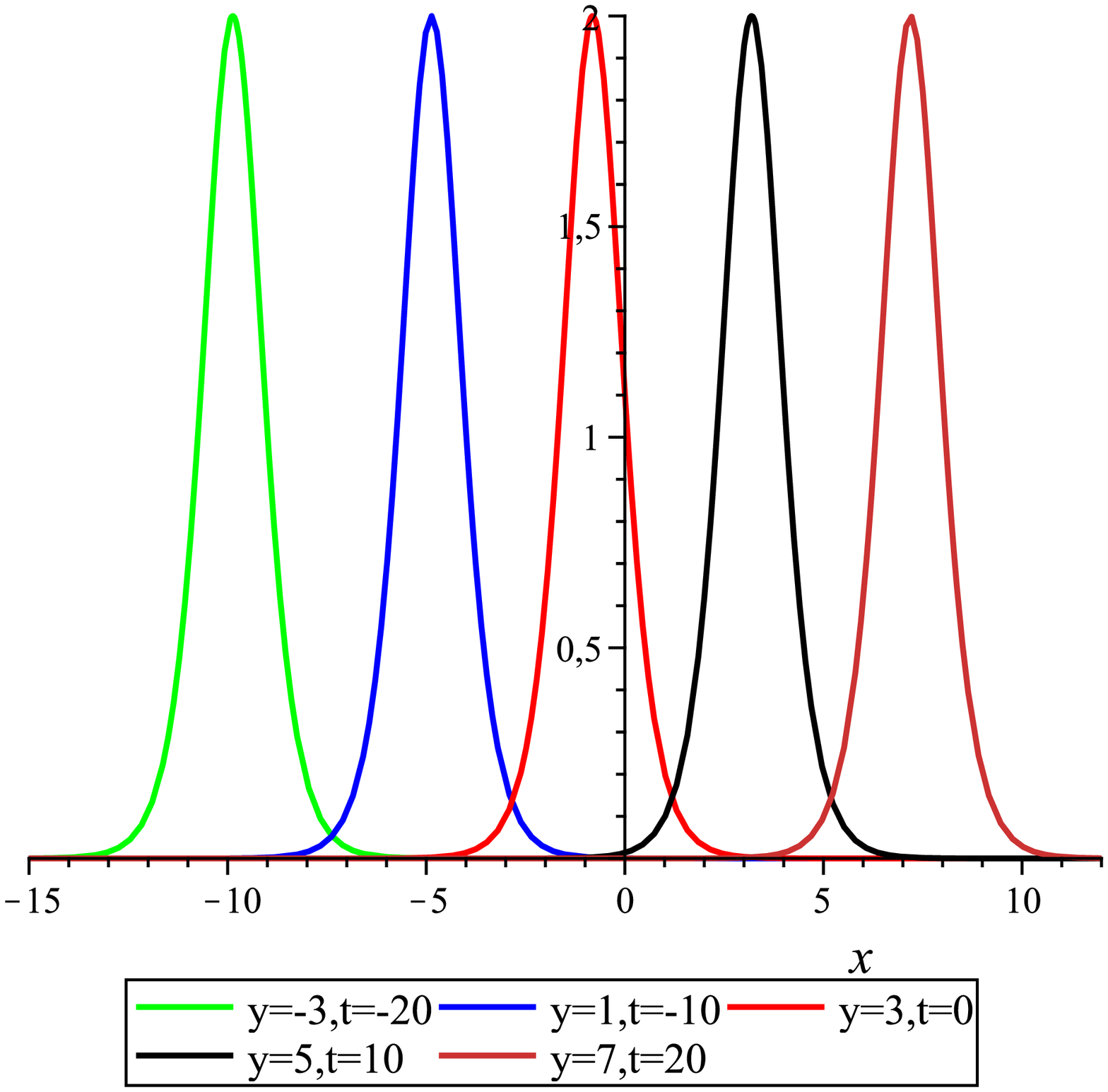}}
\caption{The movement of the waves (\ref{example2up}) as time increases.}
\end{figure}
\end{center}
\squeezeup
\noindent \textbf{(a).(ii)} $(\varepsilon_1,\varepsilon_2,\varepsilon_3)=(-1,1,1), \sigma_2=-\sigma_1$, $\sigma_1=\pm 1$.

\noindent For this case if we use Type 1 we get $k_1=0$ giving trivial solutions $u(x,y,t)=0$ and $p(x,y,t)=0$ for the nonlocal system (\ref{realnon2-a}) and (\ref{realnon2-b}).
Therefore we apply Type 2 approach based on the cross multiplication of the reduction formula. But in this case we obtain $l_1=-\bar{l}_1$ yielding
again a trivial solution.

\noindent \textbf{(a).(iii)} $(\varepsilon_1,\varepsilon_2,\varepsilon_3)=(-1,-1,1), \sigma_2=\sigma_1$, $\sigma_1=\pm 1$.

\noindent Here similar to the previous case, Type 1 approach gives a trivial solution. Using Type 2 yields the following constraints:
\begin{equation}
k_1=\bar{k}_1,\quad l_2=\bar{l}_1,\quad e^{\delta_2}=\sigma_3 e^{\delta_1},\quad e^{\delta_1+\bar{\delta}_1}=\frac{2k_1\sigma_3(l_1+\bar{l}_1)}{\rho \sigma_1},
\end{equation}
where $\sigma_3=\pm 1$. Since $k_1=\bar{k}_1$ and $a=-\bar{a}$ by (\ref{onesolcond}), we have $\omega_1=-\bar{\omega}_1$. Hence one-soliton solution of the nonlocal system (\ref{realnon3-a}) and (\ref{realnon3-b}) is given by the pair $(u(x,y,t),p(x,y,t))$ where
\begin{equation}\label{nonlocalsolrealiii}
u(x,y,t)=\frac{e^{k_1x+\omega_1t+l_1y+\delta_1}}{1+\rho \sigma_3 e^{2k_1x+(l_1+\bar{l}_1)y}}
\end{equation}
and
\begin{equation}\label{nonlocalsolrealiiip(x,y,t)}
p(x,y,t)=\frac{8k_1^2\sigma_3e^{2k_1x+(l_1+\bar{l}_1y)}}{(1+\rho\sigma_3e^{2k_1x+(l_1+\bar{l}_1y)})^2}.
\end{equation}
The above solutions are nonsingular and bounded if $\rho\sigma_3>0$. Let $a=i\beta_1$, $\omega_1=i\frac{k_1^2}{\beta_1}=i\beta_2$, $l_1=\alpha_3+i\beta_3$, where $\alpha_3  \in \mathbb{R}$, $\beta_j\in \mathbb{R}$ for $j=1,2,3$. Let also $\rho \sigma_3=e^{2\delta}$. Then from (\ref{nonlocalsolrealiii}) and
(\ref{nonlocalsolrealiiip(x,y,t)}) we have
\begin{equation}
|u(x,y,t)|^2=k_1\sigma_1\alpha_3 \mathrm{sech}^2(k_1x+\alpha_3 y+\delta) \quad \mathrm{and}\quad  p(x,y,t)=2k_1^2 \mathrm{sech}^2(k_1x+\alpha_3 y+\delta).
\end{equation}
They are bell-shaped soliton solutions.

\noindent \textbf{(b)}\, $v(x,y,t)=\rho \bar{u}(\varepsilon_1x,\varepsilon_2y,\varepsilon_3t)$, $\rho=\pm -1$, and $\varepsilon_k=\pm 1$, $k=1,2,3$.

\noindent By this nonlocal reduction we obtain seven different nonlocal reduced systems. As we noted previously the systems corresponding to (v)-(vii) are same with the systems obtained by the reduction $v(x,y,t)=\rho u(\varepsilon_1x,\varepsilon_2y,\varepsilon_3)$ and we have already presented one-soliton solutions of these systems in the previous part (a). Therefore here we will only consider one-soliton solutions of the nonlocal reduced systems corresponding to (i)-(iv).

\noindent \textbf{(b).(i)} $(\varepsilon_1,\varepsilon_2,\varepsilon_3)=(1,1,-1)$, $\sigma_2=\sigma_1$, $\sigma_1=\pm 1$.

\noindent When we use Type 1 with the reduction formula we obtain the following constraints:
\begin{equation}
k_1=\bar{k}_1,\quad l_2=\bar{l}_1,\quad e^{\delta_2}=\rho e^{\bar{\delta}_1}.
\end{equation}
In this case we have $\omega_1=-\bar{\omega}_1$. Hence one-soliton solution of the nonlocal reduced system (\ref{N=2nonlocal2-a}) and (\ref{N=2nonlocal2-b}) with
$(\varepsilon_1,\varepsilon_2,\varepsilon_3)=(1,1,-1)$ is given by the pair $(u(x,y,t), p(x,y,t))$ where
\begin{equation}\label{nonlocalsolcompi}
u(x,y,t)=\frac{e^{k_1x+\omega_1t+l_1y+\delta_1}}{1+\frac{\sigma_1}{2k_1(l_1+\bar{l}_1)}e^{2k_1x+(l_1+\bar{l}_1)y+\delta_1+\bar{\delta}_1}}
\end{equation}
and
\begin{equation}\label{nonlocalsolcompip(x,y,t)}
p(x,y,t)=\frac{4k_1\sigma_1e^{2k_1x+(l_1+\bar{l}_1y+\delta_1+\bar{\delta}_1)}}{(l_1+\bar{l}_1)[1+\frac{\sigma_1}{2k_1(l_1+\bar{l}_1)}e^{2k_1x+(l_1+\bar{l}_1)y+\delta_1
+\bar{\delta}_1}]^2}.
\end{equation}
These solutions are nonsingular and bounded for $\frac{\sigma_1}{k_1\alpha_3}>0$. Let $a=i\beta_1$, $\omega_1=i\frac{k_1^2}{\beta_1}=i\beta_2$, $l_1=\alpha_3+i\beta_3$, $e^{\delta_1}=\alpha_4+i\beta_4$, where $\alpha_3, \alpha_4 \in \mathbb{R}$, $\beta_j\in \mathbb{R}$ for $1\leq j\leq4$. Let also $\frac{\sigma_1(\alpha_4^2+\beta_4^2)}{4k_1\alpha_3}=e^{2\delta}$. Then from (\ref{nonlocalsolcompi}) and (\ref{nonlocalsolcompip(x,y,t)}) we get
\begin{equation}
|u(x,y,t)|^2=\frac{k_1\alpha_3}{\sigma_1}\mathrm{sech}^2(k_1x+\alpha_3y+\delta)\quad \mathrm{and}\quad p(x,y,t)=2k_1\mathrm{sech}^2(k_1x+\alpha_3y+\delta).
\end{equation}
These are bell-shaped soliton solutions.

\noindent \textbf{(b).(ii)} $(\varepsilon_1,\varepsilon_2,\varepsilon_3)=(1,-1,-1)$, $\sigma_2=-\sigma_1$, $\sigma_1=\pm 1$.

\noindent Using Type 1 yields the following constraints:
\begin{equation}
k_1=\bar{k}_1,\quad l_2=-\bar{l}_1,\quad e^{\delta_2}=\rho e^{\bar{\delta}_1}.
\end{equation}
Here we also get $\omega_1=-\bar{\omega}_1$ directly. Then we obtain one-soliton solution $(u(x,y,t),p(x,y,t))$ of the nonlocal reduced system (\ref{N=2nonlocal2-a}) and (\ref{N=2nonlocal2-b}) with
$(\varepsilon_1,\varepsilon_2,\varepsilon_3)=(1,-1,-1)$ as
\begin{equation}\label{nonlocalsolcompii}
u(x,y,t)=\frac{e^{k_1x+\omega_1t+l_1y+\delta_1}}{1+\frac{\sigma_1}{2k_1(l_1+\bar{l}_1)}e^{2k_1x+\delta_1+\bar{\delta}_1}\cosh((l_1+\bar{l}_1)y)}
\end{equation}
and
\begin{equation}\label{nonlocalsolcompiip(x,y,t)}
p(x,y,t)=\frac{4k_1\sigma_1e^{2k_1x+\delta_1+\bar{\delta}_1}\cosh ((l_1+\bar{l}_1)y) }{(l_1+\bar{l}_1)[1+\frac{\sigma_1}{2k_1(l_1+\bar{l}_1)}e^{2k_1x+\delta_1+\bar{\delta}_1}\cosh((l_1+\bar{l}_1)y)]^2}.
\end{equation}
Let $a=i\beta_1$, $\omega_1=i\frac{k_1^2}{\beta_1}=i\beta_2$, $l_1=\alpha_3+i\beta_3$, $e^{\delta_1}=\alpha_4+i\beta_4$, where $\alpha_3, \alpha_4 \in \mathbb{R}$, $\beta_j\in \mathbb{R}$ for $1\leq j\leq 4$. Then from (\ref{nonlocalsolcompii}) and (\ref{nonlocalsolcompiip(x,y,t)}) we obtain
\begin{equation}
|u(x,y,t)|^2=\frac{e^{2k_1x+2\alpha_3y}(\alpha_4^2+\beta_4^2)}{[1+\frac{\sigma_1(\alpha_4^2+\beta_4^2)}{4k_1\alpha_3}e^{2k_1x}\cosh(2\alpha_3y)]^2},\quad
p(x,y,t)=\frac{k_1\sigma_1(\alpha_4^2+\beta_4^2)e^{2k_1x}\cosh(2\alpha_3y)}{\alpha_3[1+\frac{\sigma_1(\alpha_4^2+\beta_4^2)}{4k_1\alpha_3}e^{2k_1x}
\cosh(2\alpha_3y)]^2}.
\end{equation}
The above solutions are nonsingular and bounded for $\frac{\sigma_1}{k_1\alpha_3}>0$. As in the nonlocal case (a).(i), the function
$|u(x,y,t)|^2$ defines a solitoff and $p(x,y,t)$ a V-type solitary wave solution.

\noindent \textbf{(b).(iii)} $(\varepsilon_1,\varepsilon_2,\varepsilon_3)=(-1,-1,-1)$, $\sigma_2=\sigma_1$, $\sigma_1=\pm 1$.

\noindent If we use Type 1, we get $k_1=-\bar{k}_1$ which gives trivial solutions $u(x,y,t)=0$ and $p(x,y,t)=0$. Therefore we apply Type 2 approach. Here
we obtain the following constraints:
\begin{equation}
l_2=l_1, \quad e^{\delta_2}=\sigma_3 e^{\bar{\delta}_1},\quad e^{\delta_1+\bar{\delta}_1}=\frac{(k_1+\bar{k}_1)(l_1+\bar{l}_1)}{\rho\sigma_1},
\end{equation}
where $\sigma_3=\pm 1$. Hence one-soliton solution of the nonlocal reduced system (\ref{N=2nonlocal2-a}) and (\ref{N=2nonlocal2-b}) with
 $(\varepsilon_1,\varepsilon_2,\varepsilon_3)=(-1,-1,-1)$ is obtained as the pair $(u(x,y,t),p(x,y,t))$ where
\begin{equation}\label{nonlocalsolcompiii}
u(x,y,t)=\frac{e^{k_1x+\omega_1t+l_1y+\delta_1}}{1+\rho e^{(k_1+\bar{k}_1)x+(\omega_1+\bar{\omega}_1)t+(l_1+\bar{l}_1)y}}
 \end{equation}
and
\begin{equation}\label{nonlocalsolcompiiip(x,y,t)}
p(x,y,t)=\frac{2\rho(k_1+\bar{k}_1)^2e^{(k_1+\bar{k}_1)x+(\omega_1+\bar{\omega}_1)t+(l_1+\bar{l}_1)y}}{[1+\rho e^{(k_1+\bar{k}_1)x+(\omega_1+\bar{\omega}_1)t+(l_1+\bar{l}_1)y}  ]^2}.
\end{equation}
 The above solutions are nonsingular and bounded if $\rho=1$. For $\rho=1$, let $k_1=\alpha_1+i\beta_1$, $\omega_1=\alpha_2+i\beta_2$, $a=i\beta_3$, $l_1=\alpha_4+i\beta_4$, where $\alpha_1, \alpha_2, \alpha_4 \in \mathbb{R}$, $\beta_j\in \mathbb{R}$ for $1\leq j\leq 4$. Here $\alpha_2=-\frac{2\alpha_1\beta_1}{\beta_3}$ and $\beta_2=\frac{(\alpha_1^2-\beta_1^2)}{\beta_3}$. Then from (\ref{nonlocalsolcompiii}) and (\ref{nonlocalsolcompiiip(x,y,t)}) we get
 \begin{equation}\label{nonlocalsolcompiiimodulus}
 |u(x,y,t)|^2=\alpha_1\alpha_4 \sigma_1 \mathrm{sech}^2(\alpha_1x+\alpha_2t+\alpha_4y+\delta),\quad p(x,y,t)=2\alpha_1^2\mathrm{sech}^2(\alpha_1x+\alpha_2t+\alpha_4y+\delta).
 \end{equation}
 The solutions are nonsingular and bounded bell-shaped soliton solutions.

 \noindent \textbf{(b).(iv)} $(\varepsilon_1,\varepsilon_2,\varepsilon_3)=(-1,1,-1)$, $\sigma_2=-\sigma_1$, $\sigma_1=\pm 1$.

\noindent For this case if we use Type 1 we get $k_1=-\bar{k}_1$ which gives trivial solutions $u(x,y,t)=0$ and $p(x,y,t)=0$. We then apply Type 2 and get $l_1=-\bar{l}_1$ which also yields trivial solutions.

\subsection{Two-soliton solutions of the local reduced systems}

Similar to one-soliton solution, we only consider the local reduction $v(x,y,t)=\rho u(x,y,t)$ since the second local reduction $v(x,y,t)=\rho \bar{u}(x,y,t)$ is consistent if $a=\bar{a}$ different than the constraint $a=-\bar{a}$ that we obtain for two-soliton solution of Maccari system.

\noindent \textbf{(a)}\, $v(x,y,t)=\rho u(x,y,t)$, $\rho$ is a real constant.

\noindent Under this reduction, when we use Type 1 approach with two-soliton solution (\ref{twosoliton}), in addition to the conditions (\ref{constraintstwo}) we get the following constraints:
\begin{equation}
s_j=l_j,\quad e^{\alpha_j}=\rho e^{\delta_j}, \quad j=1, 2.
\end{equation}
\noindent \textbf{Example 3.} Let us take the parameters of two-soliton solution as $(k_1,k_2,l_1,l_2,a)=(2,\frac{1}{2},\frac{1}{3},1,2i)$ with $\sigma_1=\sigma_2=\rho=e^{\delta_1}=e^{\delta_2}=1$. Then the graphs of the two-soliton solutions of the local reduced $2$-component Maccari system (\ref{N=2local1-a}) and (\ref{N=2local1-b}) at $t=0$ are given in Figure 5.
\begin{center}
\begin{figure}[h!]
\centering
\subfloat[]{\includegraphics[width=0.35\textwidth]{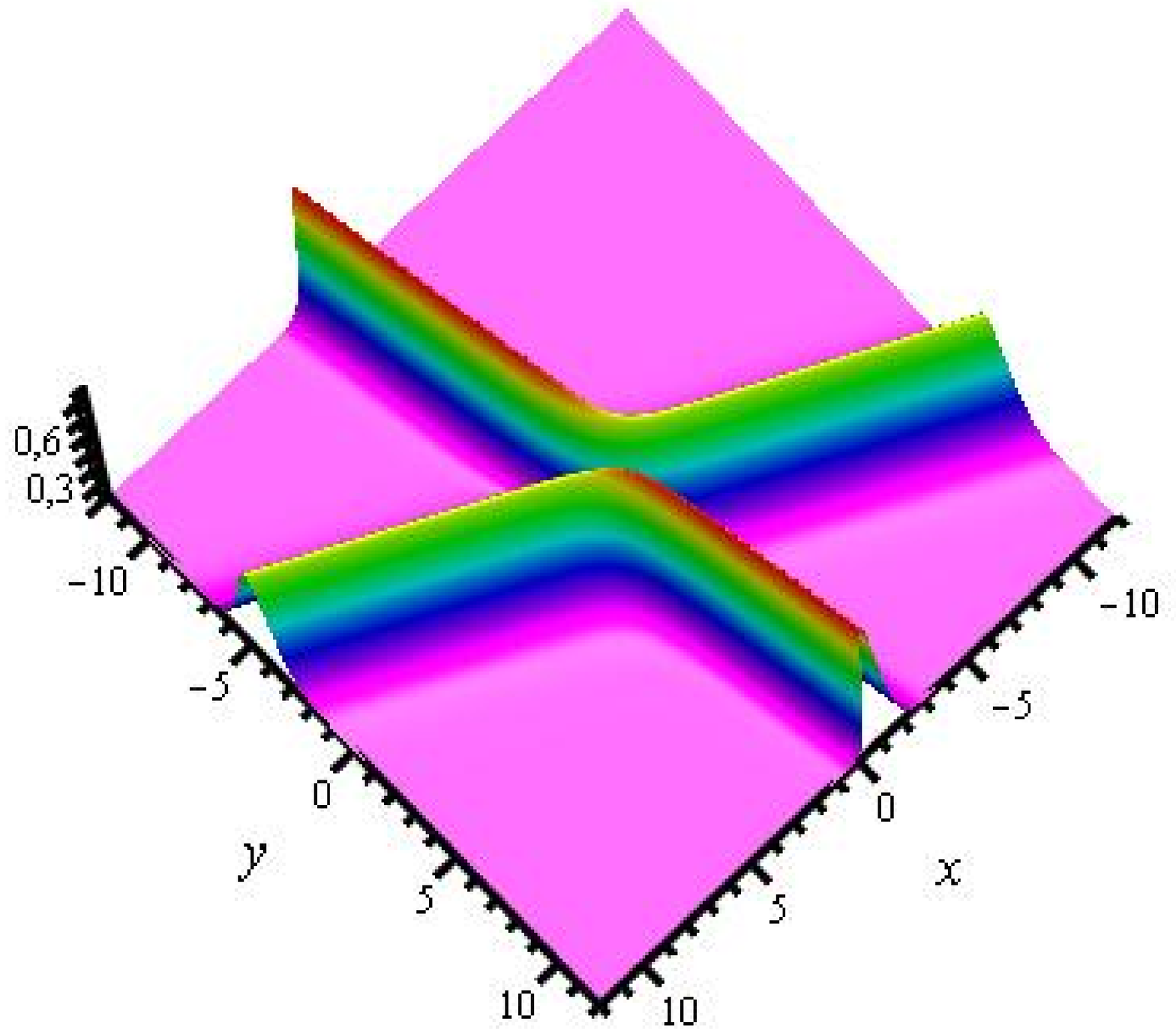}}\hspace{2cm}
\subfloat[] {\includegraphics[width=0.35\textwidth]{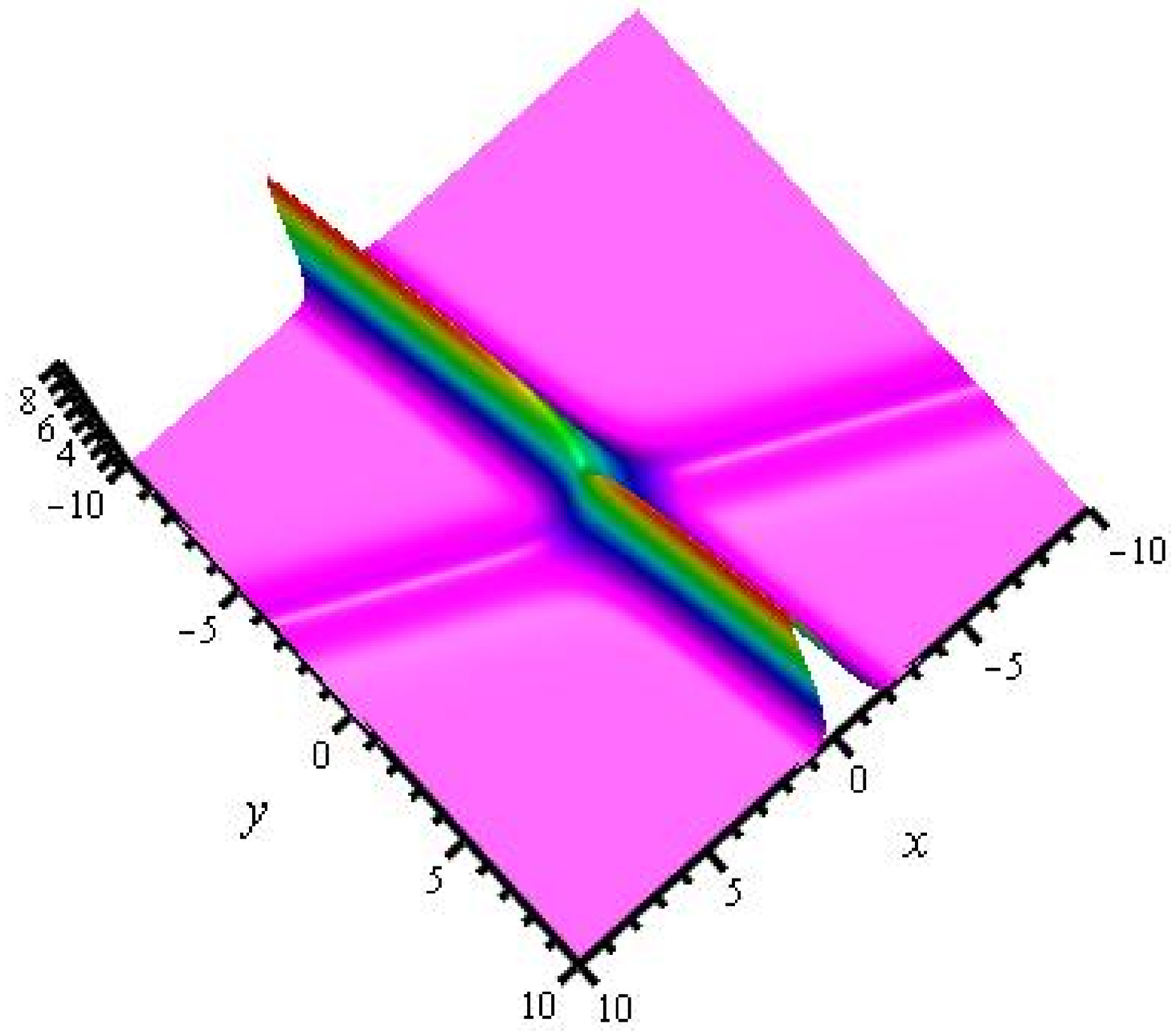}}
\caption{Two-soliton solutions of the local system (\ref{N=2local1-a}) and (\ref{N=2local1-b}) at $t=0$ for the parameters  $(k_1,k_2,l_1,l_2,a)=(2,\frac{1}{2},\frac{1}{3},1,2i)$ with $\sigma_1=\sigma_2=\rho=e^{\delta_1}=e^{\delta_2}=1$. (a) Two V-type solitary waves from $|u(x,y,t)|^2$, (b) V-type interaction of localized solitons from $p(x,y,t)$.}
\end{figure}
\end{center}
\squeezeup
Figures 6(a) and 6(b) show the interactions of the waves from $|u(x,y,t)|^2$ and $p(x,y,t)$ for different times in different positions at the $y$-axis, respectively.
\begin{center}
\begin{figure}[h!]
\centering
\subfloat[]{\includegraphics[width=0.30\textwidth]{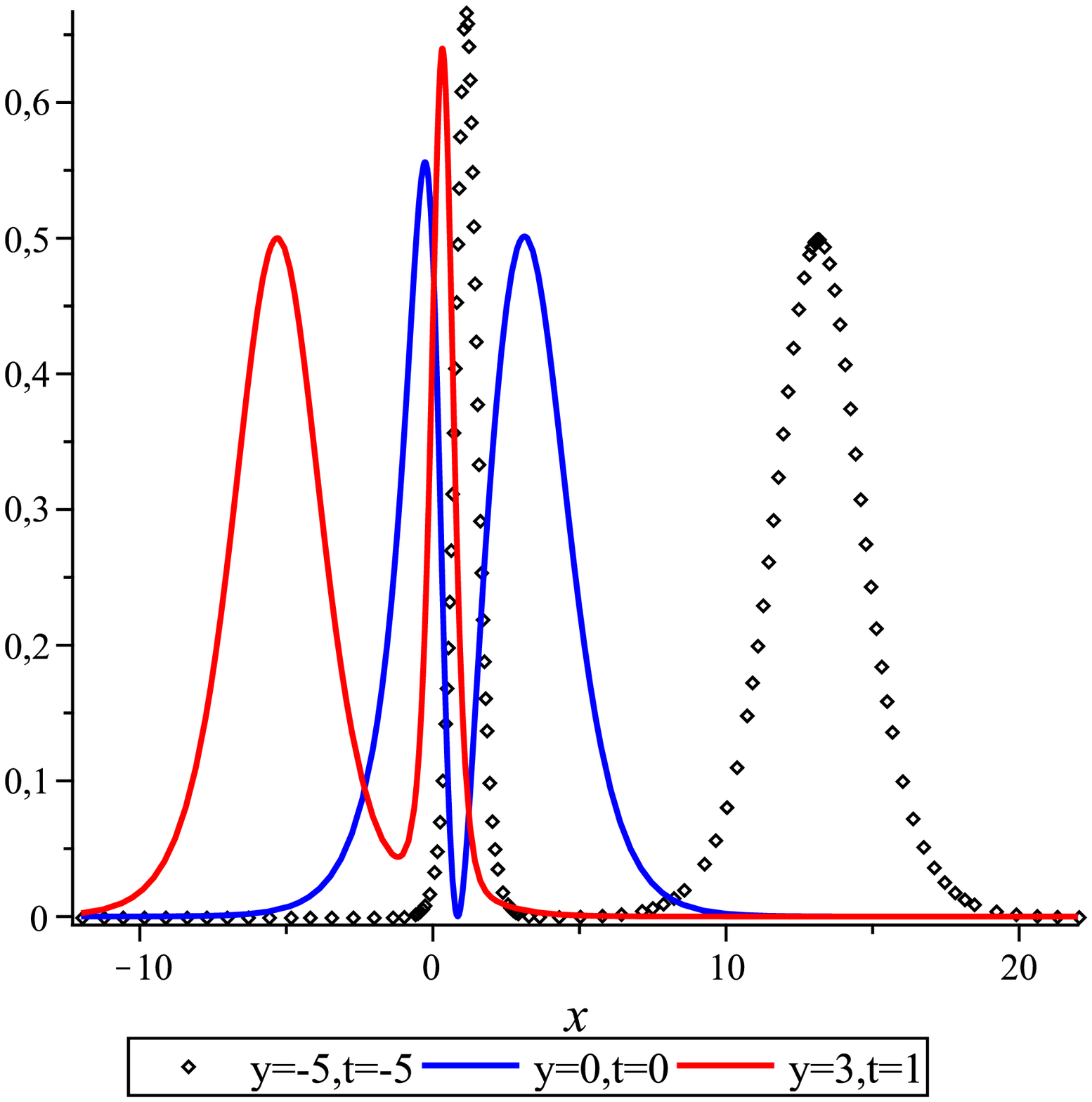}}\hspace{2cm}
\subfloat[] {\includegraphics[width=0.30\textwidth]{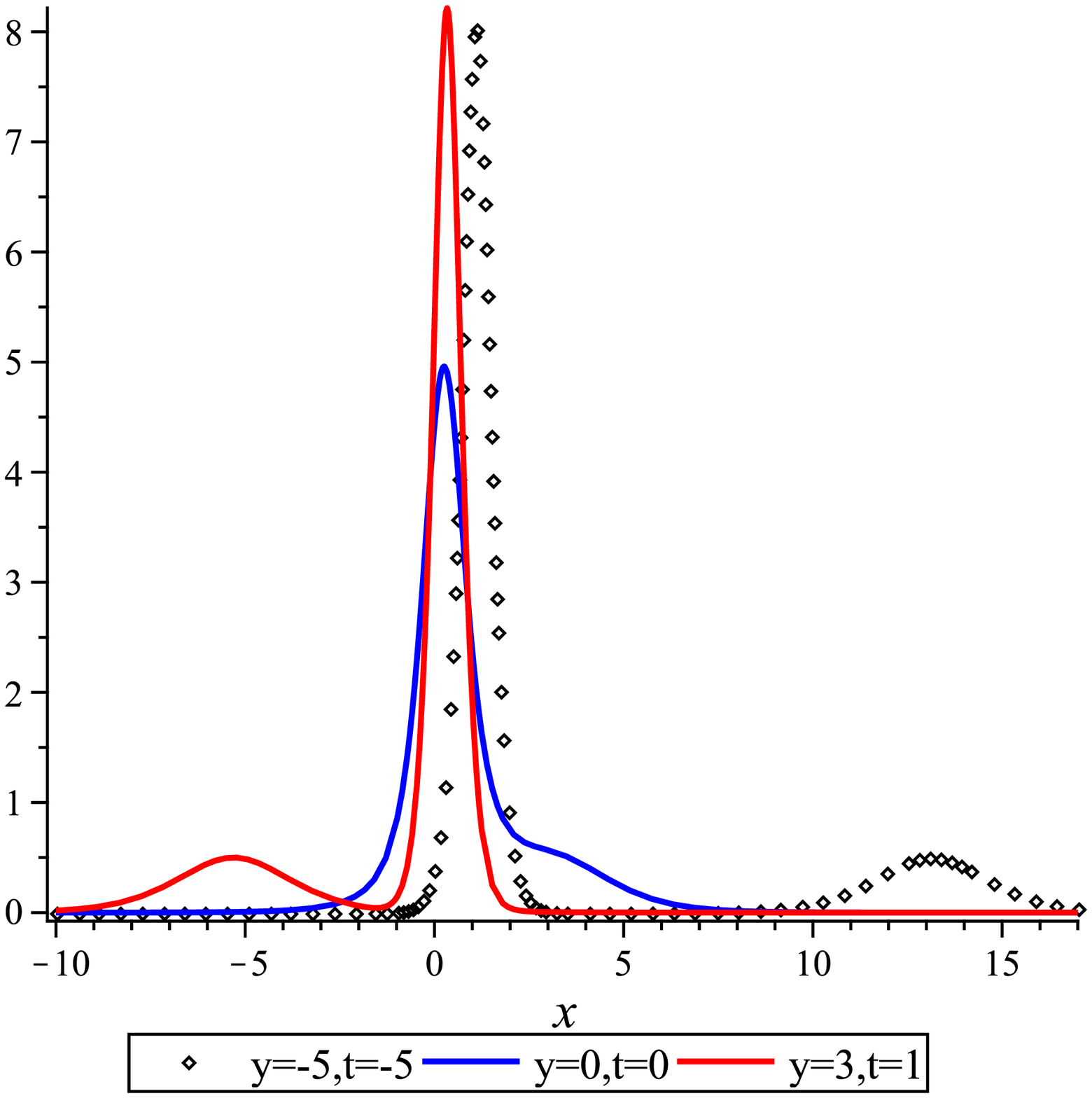}}
\caption{The movement of the waves as time increases.}
\end{figure}
\end{center}
\squeezeup
\subsection{Two-soliton solutions of the nonlocal reduced systems}

In this section we obtain two-soliton solutions of nonlocal systems obtained from $3$-component Maccari systems by two nonlocal reductions.

\noindent \textbf{(a)}\, $v(x,y,t)=\rho u(\varepsilon_1x,\varepsilon_2y,\varepsilon_3t)$, $\rho=\pm 1$, and $\varepsilon_j=\pm 1$, $j=1,2,3$.

Under this reduction we will consider soliton solutions of the three different S-reversal nonlocal reduced $2$-component Maccari systems.

\newpage

\noindent \textbf{(a).(i)} $(\varepsilon_1,\varepsilon_2,\varepsilon_3)=(1,-1,1), \sigma_2=-\sigma_1$, $\sigma_1=\pm 1$.

If we use this nonlocal reduction with two-soliton solution (\ref{twosoliton}) and apply Type 1, besides (\ref{constraintstwo}) we get the following constraints:
\begin{equation}
s_j=-l_j,\quad e^{\alpha_j}=\rho e^{\delta_j}, \quad j=1, 2.
\end{equation}
\noindent \textbf{Example 4.} Let us choose the two-soliton solution parameters as $(k_1,k_2,l_1,l_2,a)=(1+\frac{i}{2},\frac{1}{4},\frac{1}{3},1,i)$
 with $\sigma_1=e^{\delta_1}=e^{\delta_2}=\rho=1$. Then the graphs of the two-soliton solutions of (\ref{realnon1-a}) and (\ref{realnon1-b}) are given in
 Figure 7.
 \begin{center}
\begin{figure}[h!]
\centering
\subfloat[]{\includegraphics[width=0.35\textwidth]{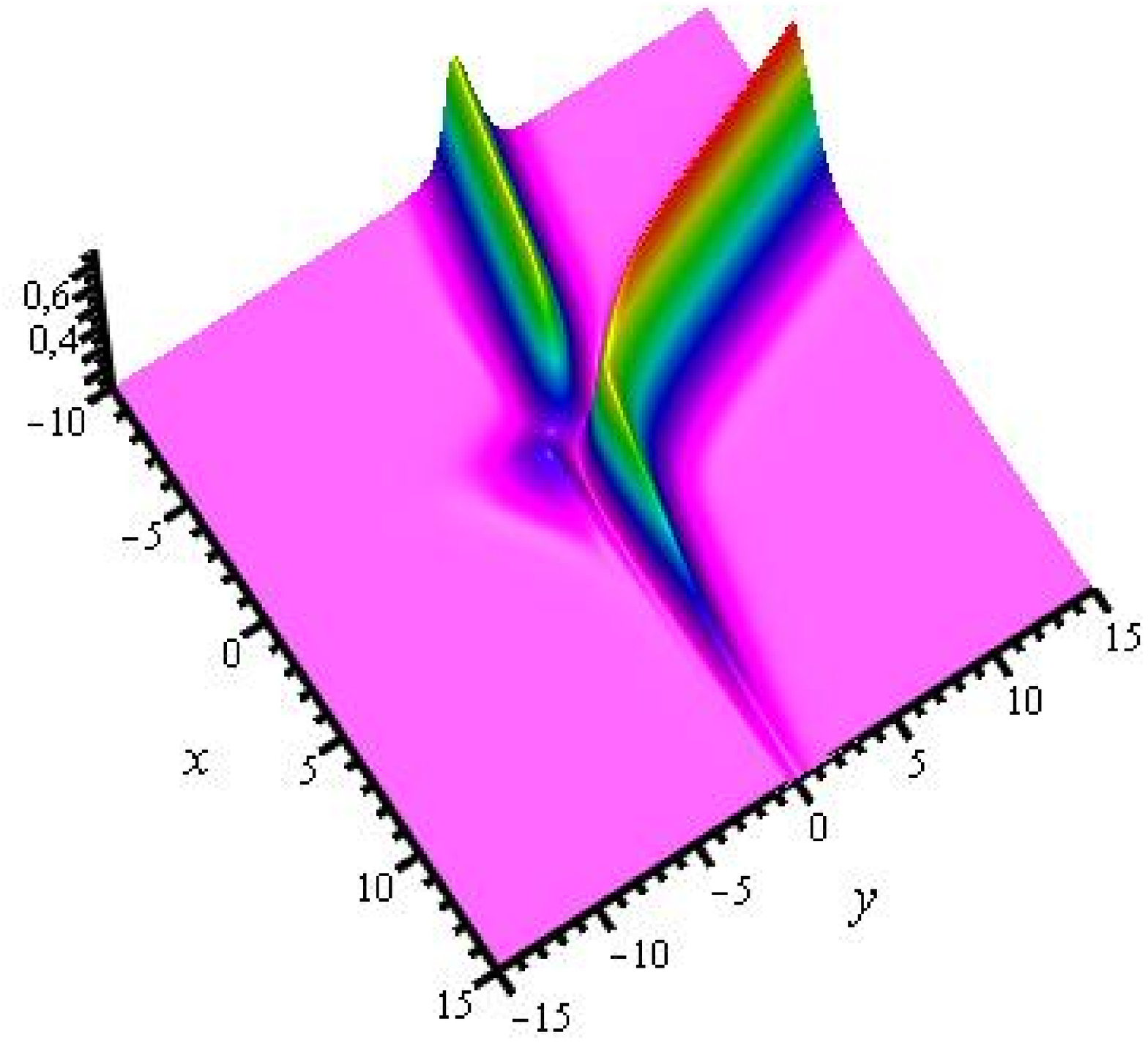}}\hspace{2cm}
\subfloat[] {\includegraphics[width=0.35\textwidth]{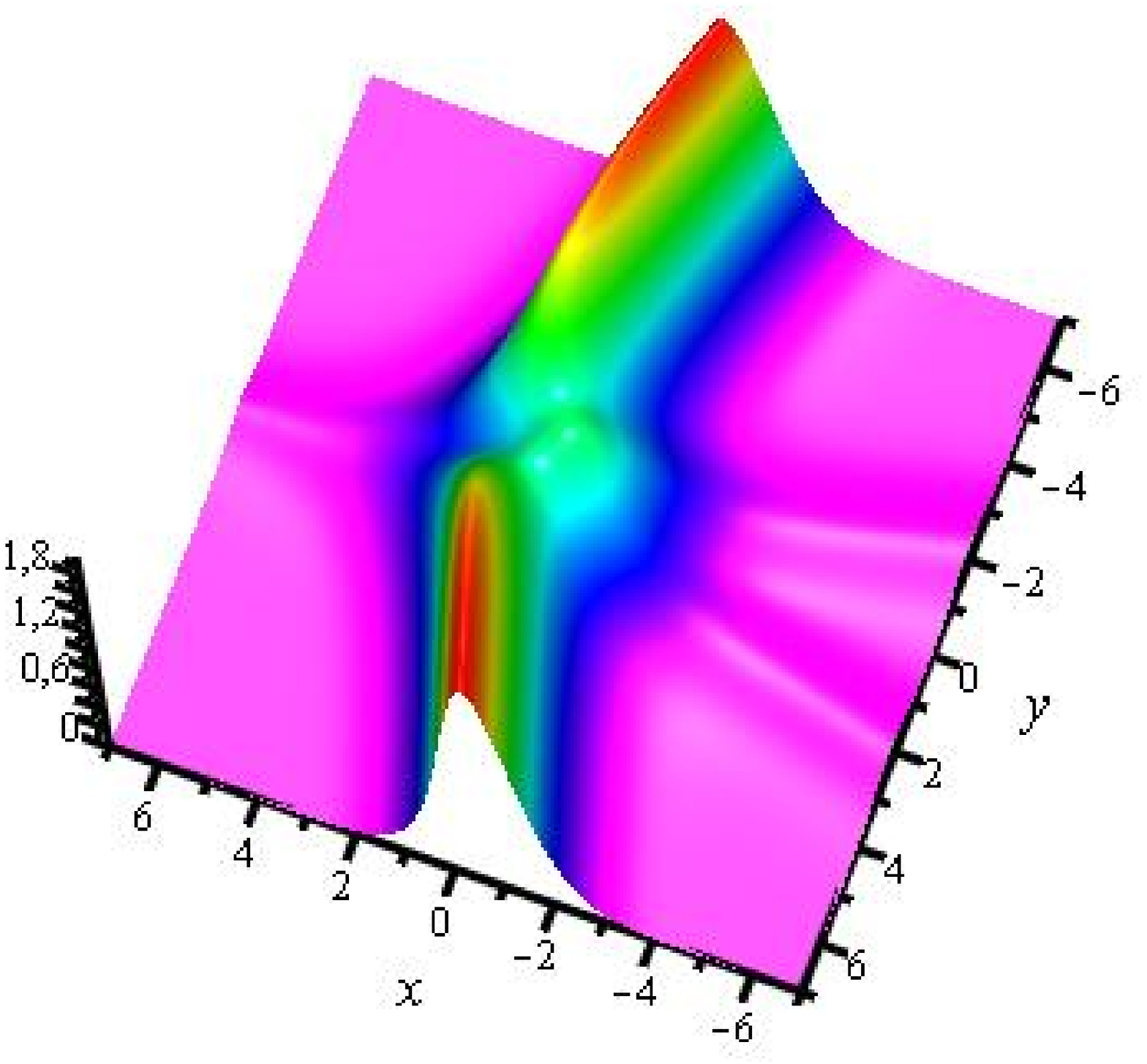}}
\caption{Two-soliton solutions of the nonlocal system (\ref{realnon1-a}) and (\ref{realnon1-b}) at $t=0$ for the parameters  $(k_1,k_2,l_1,l_2,a)=(1+\frac{i}{2},\frac{1}{4},\frac{1}{3},1,i)$ with $\sigma_1=e^{\delta_1}=e^{\delta_2}=\rho=1$. (a) V-type interaction of two solitoffs from $|u(x,y,t)|^2$, (b) Interaction of V-type solitary waves from $p(x,y,t)$.}
\end{figure}
\end{center}
\squeezeup
\newpage
Figures 8(a) and 8(b) depict the interactions of the waves from $|u(x,y,t)|^2$ and $p(x,y,t)$ for different times in different positions at the $y$-axis, respectively.
\begin{center}
\begin{figure}[h!]
\centering
\subfloat[]{\includegraphics[width=0.29\textwidth]{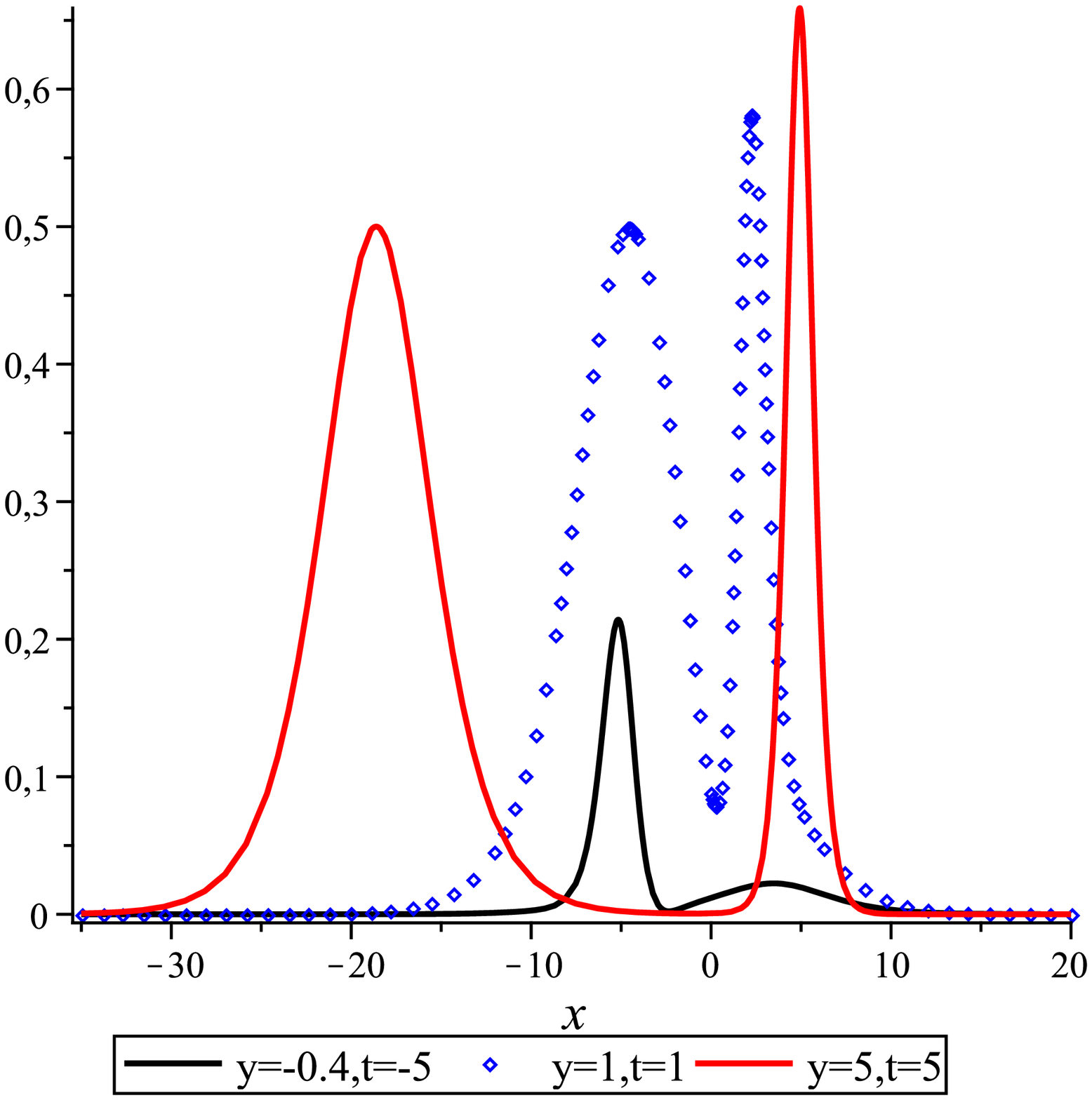}}\hspace{2cm}
\subfloat[] {\includegraphics[width=0.29\textwidth]{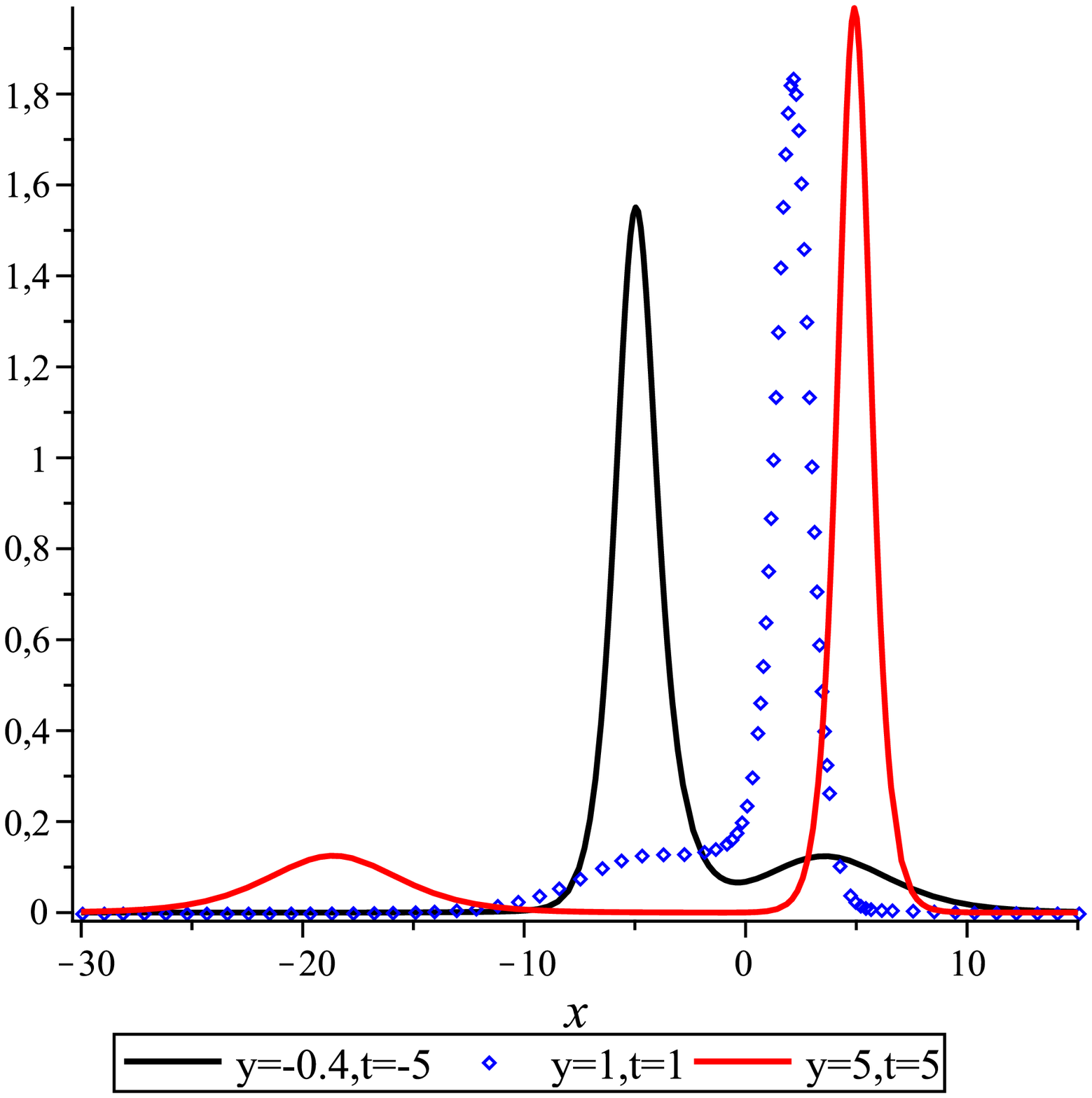}}
\caption{The movement of the waves as time increases.}
\end{figure}
\end{center}
\squeezeup
\noindent \textbf{(a).(ii)} $(\varepsilon_1,\varepsilon_2,\varepsilon_3)=(-1,1,1), \sigma_2=-\sigma_1$, $\sigma_1=\pm 1$.

When we use this nonlocal reduction with (\ref{twosoliton}) and apply Type 1,  in addition to (\ref{constraintstwo}) we obtain the following constraints:
\begin{equation}
k_2=-k_1, \quad s_1=l_2,\quad s_2=l_1,\quad e^{\alpha_1}=\rho e^{\delta_2}, \quad  e^{\alpha_2}=\rho e^{\delta_1}.
\end{equation}
\noindent  \textbf{Example 5.} Consider the two-soliton solution parameters as $(k_1, l_1,l_2,a)=(1+\frac{i}{2},\frac{1}{2},-1,2i)$
 with $\sigma_1=e^{\delta_1}=e^{\delta_2}=\rho=1$. Then the graphs of the two-soliton solutions of the nonlocal system (\ref{realnon2-a}) and (\ref{realnon2-b}) are given in
 Figure 9.
 \begin{center}
\begin{figure}[h!]
\centering
\subfloat[]{\includegraphics[width=0.35\textwidth]{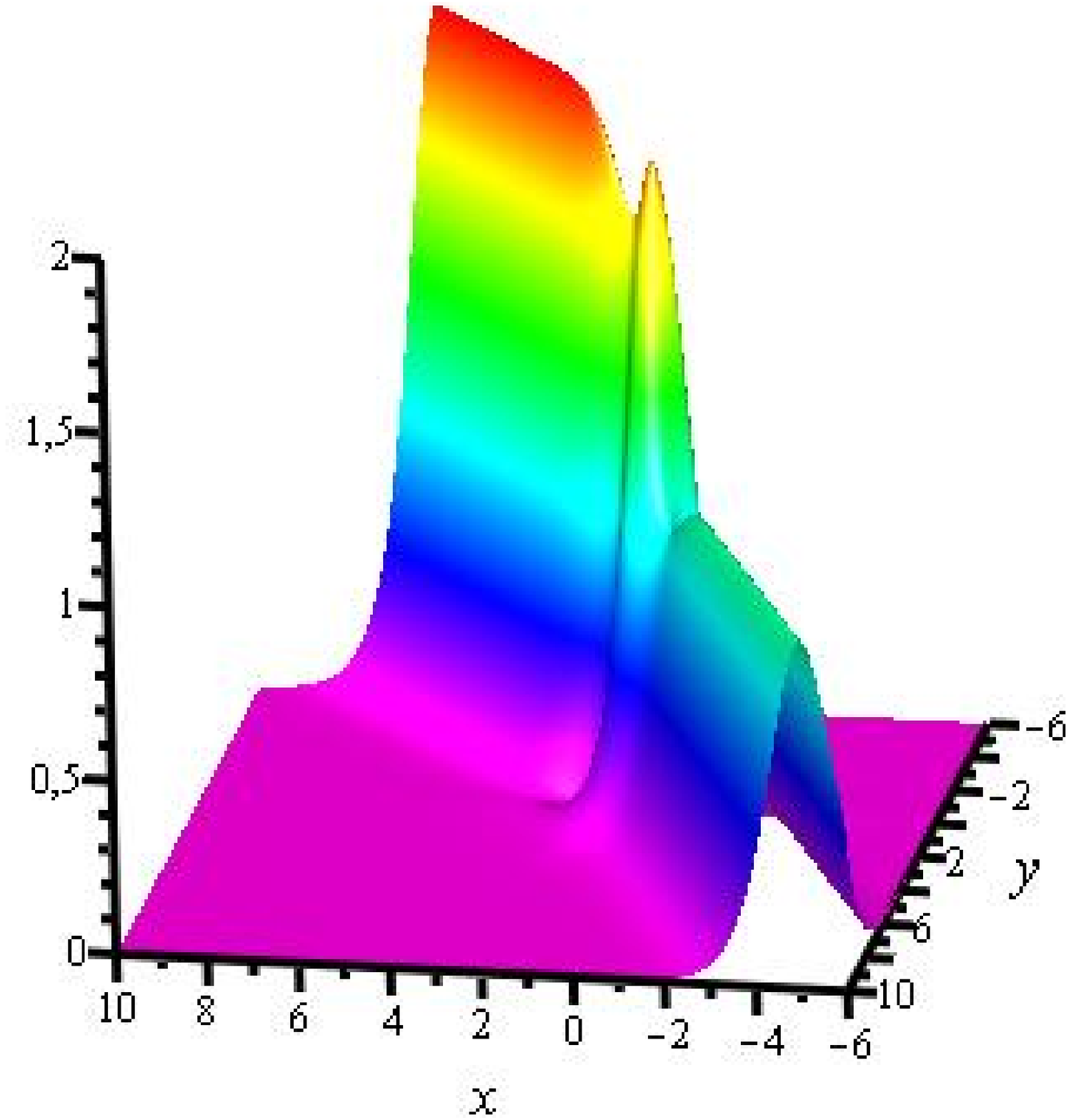}}\hspace{2cm}
\subfloat[] {\includegraphics[width=0.35\textwidth]{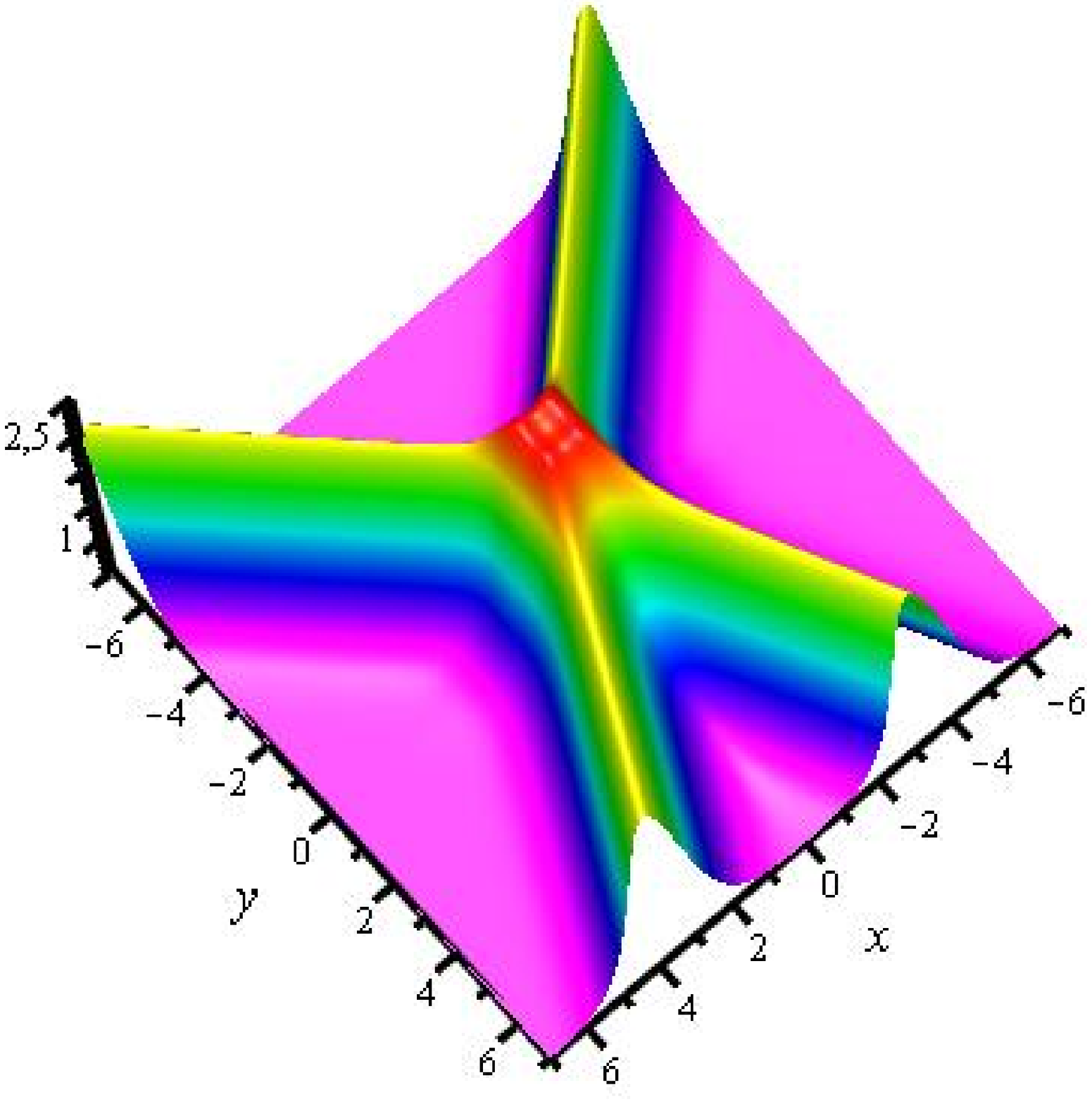}}
\caption{Two-soliton solutions of the nonlocal system (\ref{realnon2-a}) and (\ref{realnon2-b}) at $t=0$ for the parameters  $(k_1, l_1,l_2,a)=(1+\frac{i}{2},\frac{1}{2},-1,2i)$
 with $\sigma_1=e^{\delta_1}=e^{\delta_2}=\rho=1$. (a) Interaction of two solitoffs from $|u(x,y,t)|^2$, (b) Interaction of two V-type solitary waves from $p(x,y,t)$.}
\end{figure}
\end{center}
\squeezeup
\newpage
Figures 10(a) and 10(b) depict the interactions of the waves from $|u(x,y,t)|^2$ and $p(x,y,t)$ for different times in different positions at the $y$-axis, respectively.
\begin{center}
\begin{figure}[h!]
\centering
\subfloat[]{\includegraphics[width=0.29\textwidth]{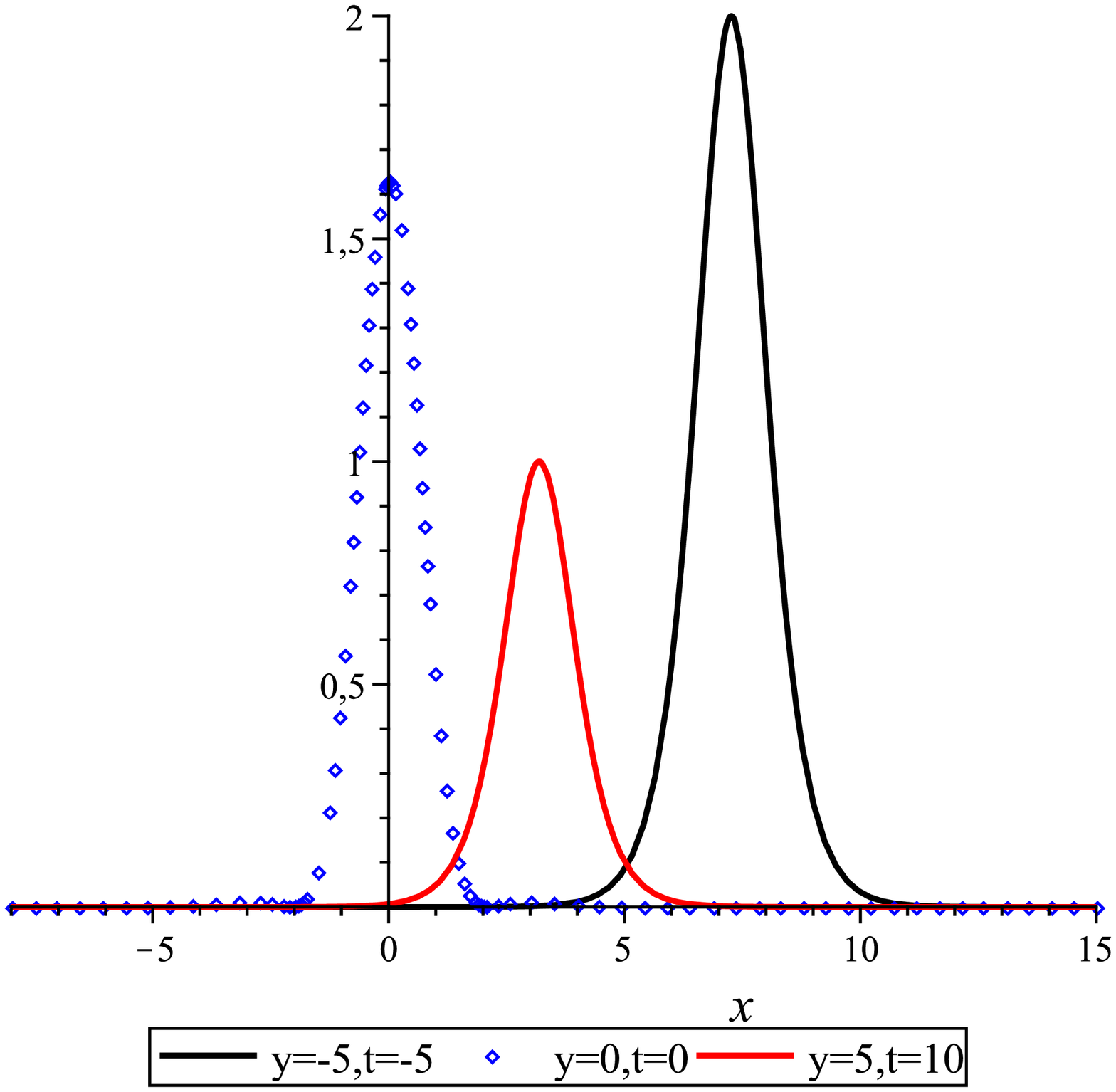}}\hspace{2cm}
\subfloat[] {\includegraphics[width=0.29\textwidth]{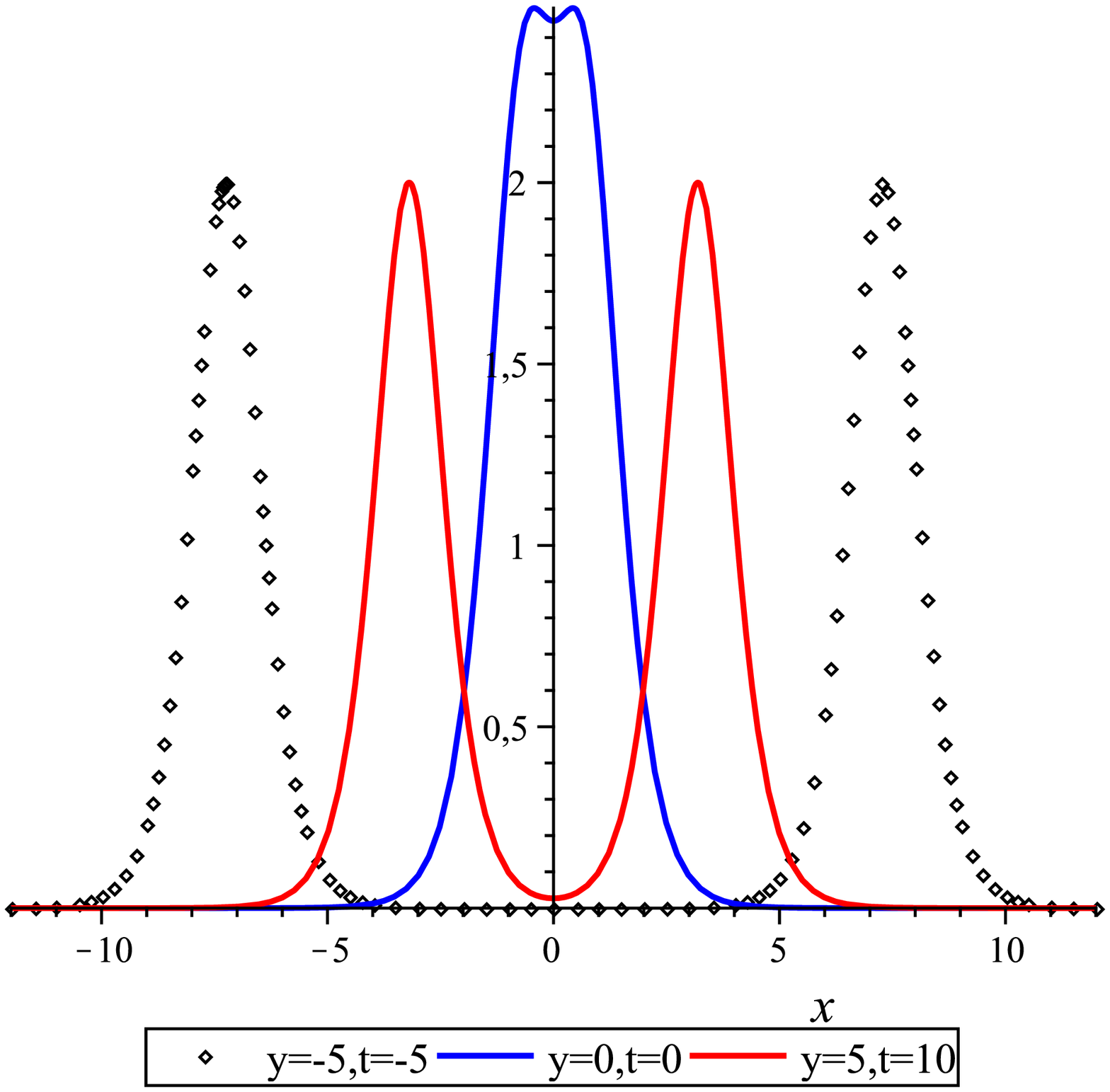}}
\caption{The movement of the waves as time increases.}
\end{figure}
\end{center}
\squeezeup
\noindent \textbf{(a).(iii)} $(\varepsilon_1,\varepsilon_2,\varepsilon_3)=(-1,-1,1), \sigma_2=\sigma_1$, $\sigma_1=\pm 1$.

For this case, applying Type 1 approach yields the conditions,
\begin{equation}
k_2=-k_1, \quad s_1=-l_2,\quad s_2=-l_1,\quad e^{\alpha_1}=\rho e^{\delta_2}, \quad  e^{\alpha_2}=\rho e^{\delta_1},
\end{equation}
besides the constraints (\ref{constraintstwo}). Here we obtain nonsingular solutions very similar to the cases (a).(i) and (a).(ii), hence we will not
give a particular example for this case.\\

\noindent \textbf{(b)}\, $v(x,y,t)=\rho \bar{u}(\varepsilon_1x,\varepsilon_2y,\varepsilon_3t)$, $\rho=\pm -1$, and $\varepsilon_j=\pm 1$, $j=1,2,3$.

In this part we will consider soliton solutions of four different nonlocal reduced $2$-component Maccari systems.

\noindent \textbf{(b).(i)} $(\varepsilon_1,\varepsilon_2,\varepsilon_3)=(1,1,-1)$, $\sigma_2=\sigma_1$, $\sigma_1=\pm 1$.

Here using two-soliton solution (\ref{twosoliton}) with this reductions and applying Type 1 yield the following constraints:
\begin{equation}
k_j=\bar{k}_j, \quad s_j=\bar{l}_j,\quad e^{\alpha_j}=\rho e^{\bar{\delta}_j},\quad j=1, 2.
\end{equation}

\noindent  \textbf{Example 6.} Consider the two-soliton solution parameters as $(k_1, k_2, l_1,l_2,a)=(1,2,\frac{1}{2},2,i)$
 with $\sigma_1=e^{\delta_1}=e^{\delta_2}=\rho=1$. Then the graphs of the two-soliton solutions of the nonlocal system (\ref{realnon2-a}) and (\ref{realnon2-b}) are given in
 Figure 11.
 \begin{center}
\begin{figure}[h!]
\centering
\subfloat[]{\includegraphics[width=0.35\textwidth]{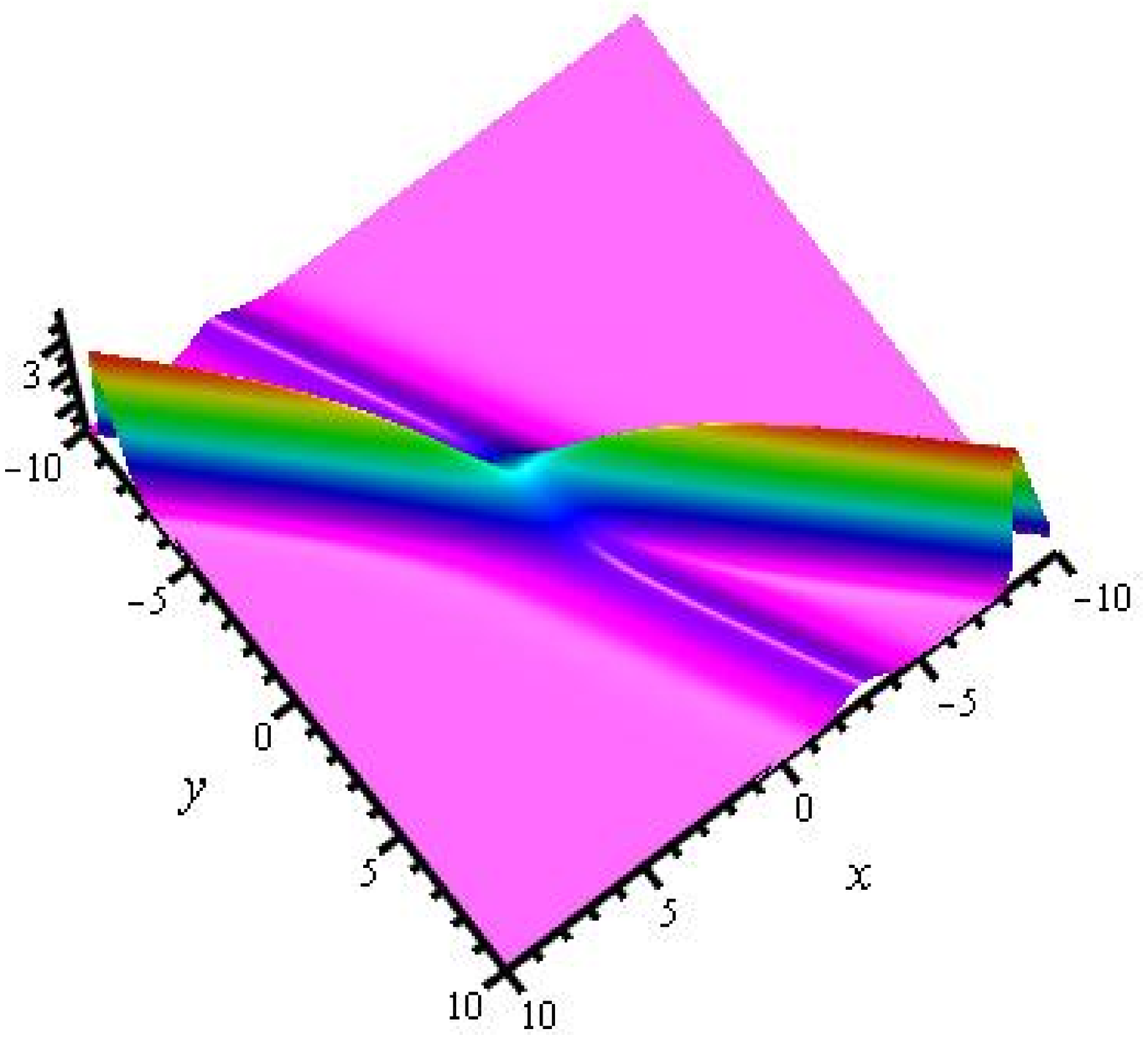}}\hspace{2cm}
\subfloat[] {\includegraphics[width=0.35\textwidth]{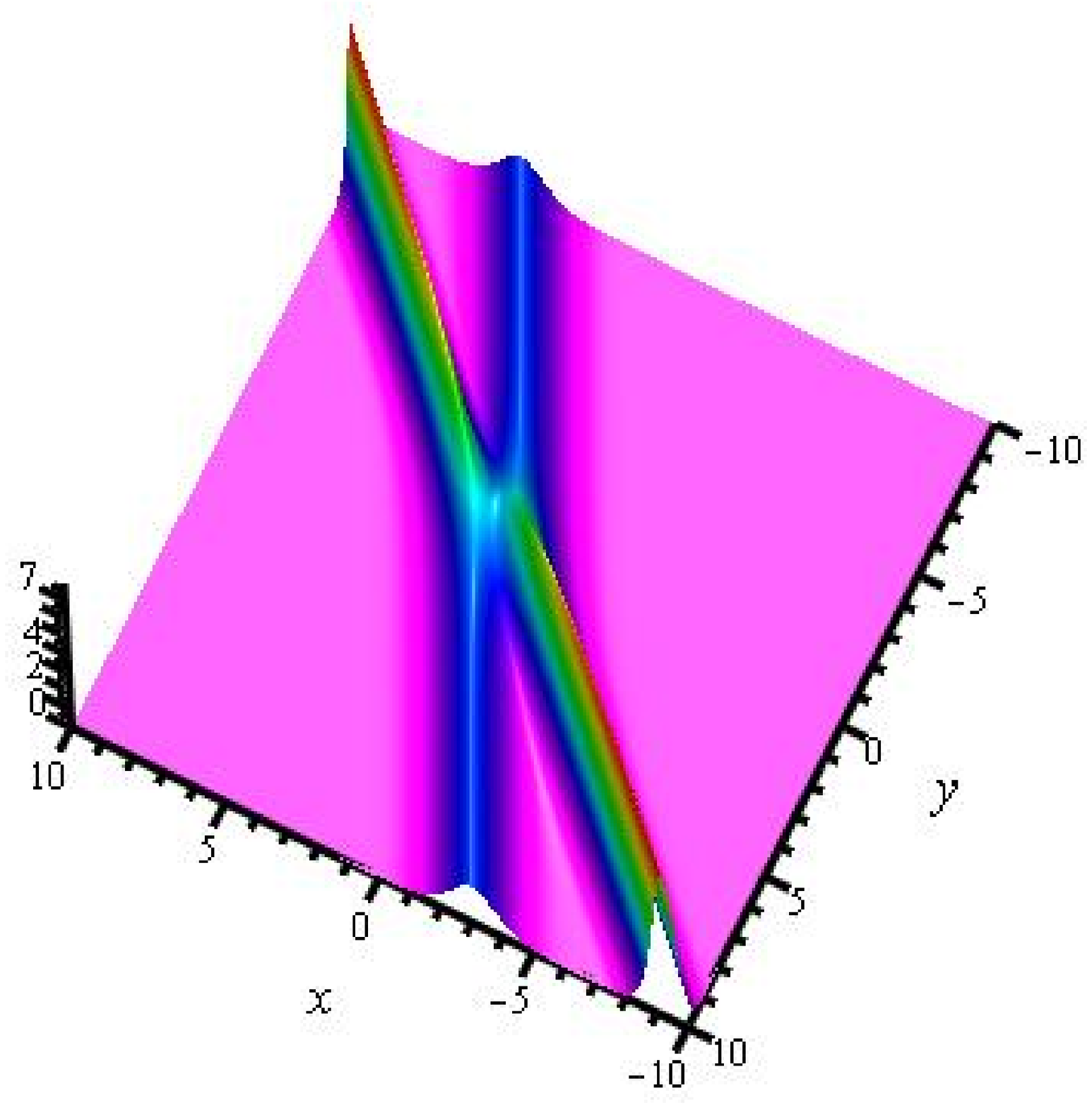}}
\caption{Two-soliton solutions of the nonlocal system (\ref{realnon2-a}) and (\ref{realnon2-b}) at $t=0$ for the parameters  $(k_1, k_2, l_1,l_2,a)=(1,2,\frac{1}{2},2,i)$
 with $\sigma_1=e^{\delta_1}=e^{\delta_2}=\rho=1$. (a) Interaction of localized two-solitons from $|u(x,y,t)|^2$, (b) Interaction of localized two-solitons from $p(x,y,t)$.}
\end{figure}
\end{center}
\squeezeup

Here the form of the waves changes as time changes but the their locations does not change. Figures 12(a) and 12(b) show the interactions of the waves from $|u(x,y,t)|^2$ and $p(x,y,t)$ for different times in different positions at the $y$-axis, respectively.
\begin{center}
\begin{figure}[h!]
\centering
\subfloat[]{\includegraphics[width=0.29\textwidth]{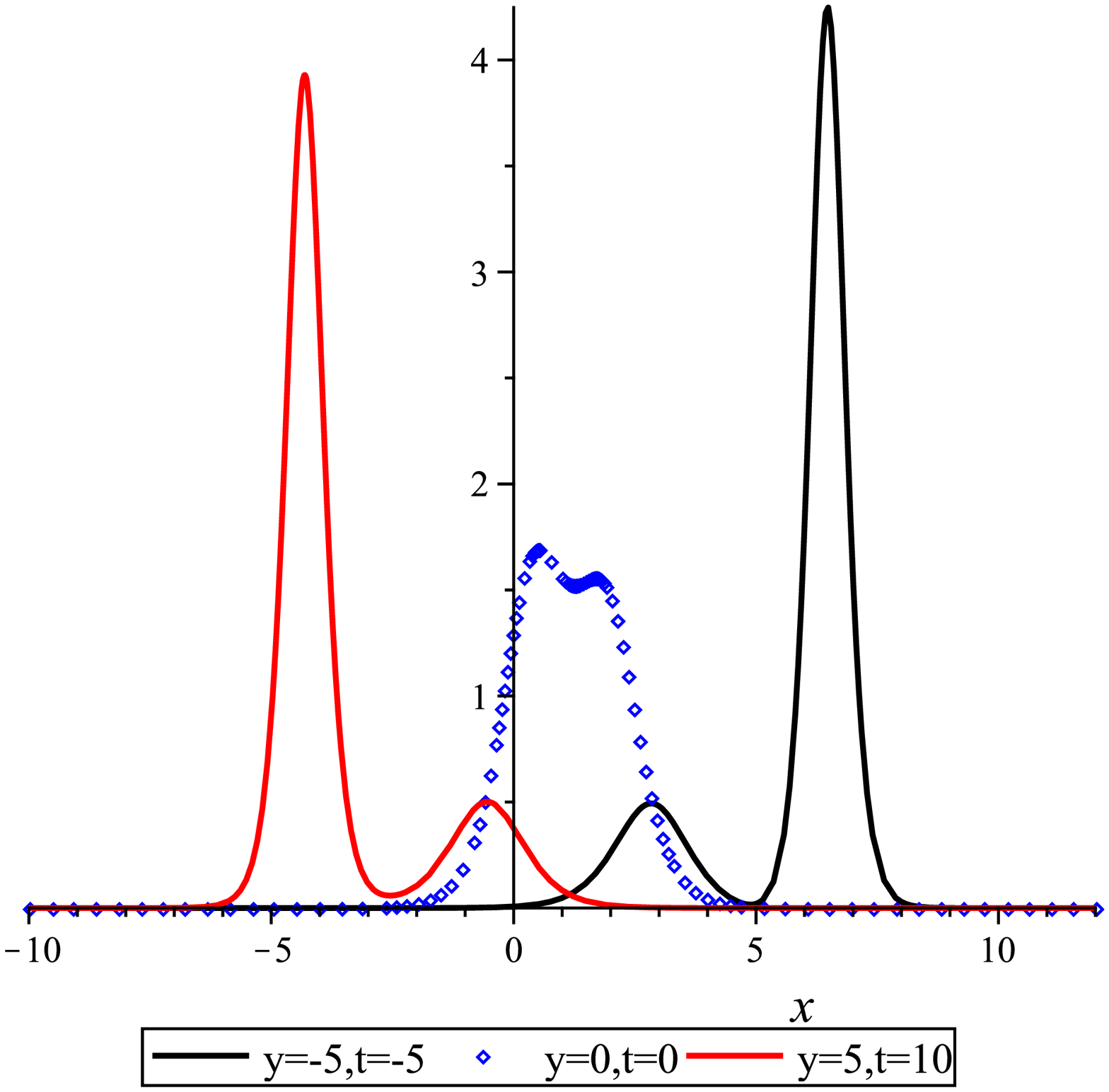}}\hspace{2cm}
\subfloat[] {\includegraphics[width=0.29\textwidth]{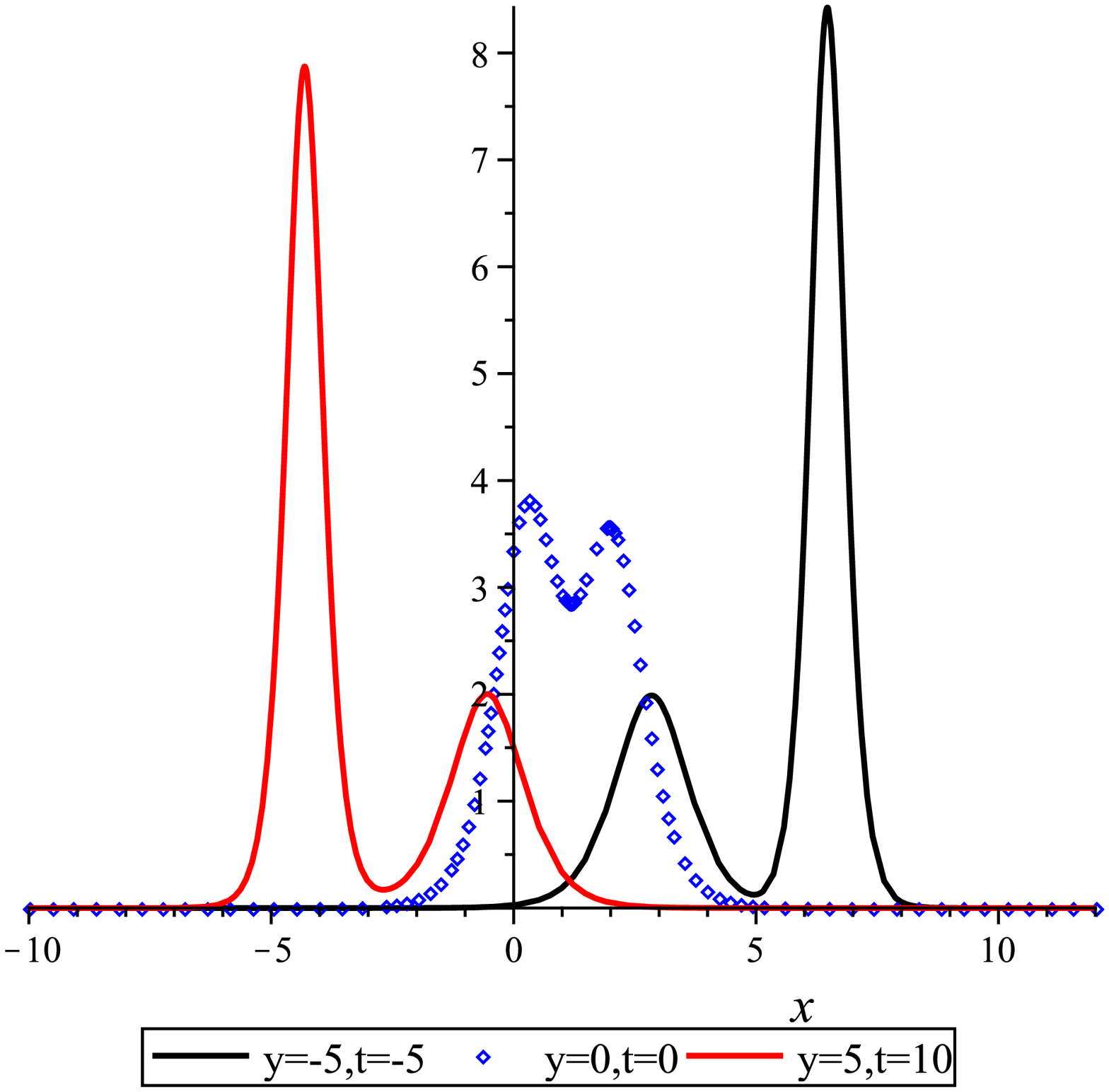}}
\caption{The movement of the waves as time increases.}
\end{figure}
\end{center}
\squeezeup
\noindent \textbf{(b).(ii)} $(\varepsilon_1,\varepsilon_2,\varepsilon_3)=(1,-1,-1)$, $\sigma_2=-\sigma_1$, $\sigma_1=\pm 1$.

\noindent Applying Type 1 approach, here we obtain the constraints,
\begin{equation}
k_j=\bar{k}_j, \quad s_j=-\bar{l}_j,\quad e^{\alpha_j}=\rho e^{\bar{\delta}_j},\quad j=1, 2.
\end{equation}

\noindent  \textbf{Example 7.} Take the two-soliton solution parameters as $(k_1, k_2, l_1,l_2,a)=(1,2,\frac{1}{2},1,i)$
 with $\sigma_1=e^{\delta_1}=e^{\delta_2}=\rho=1$. Then the graphs of the two-soliton solutions of the nonlocal system (\ref{realnon2-a}) and (\ref{realnon2-b}) are given in
 Figure 13.
 \begin{center}
\begin{figure}[h!]
\centering
\subfloat[]{\includegraphics[width=0.35\textwidth]{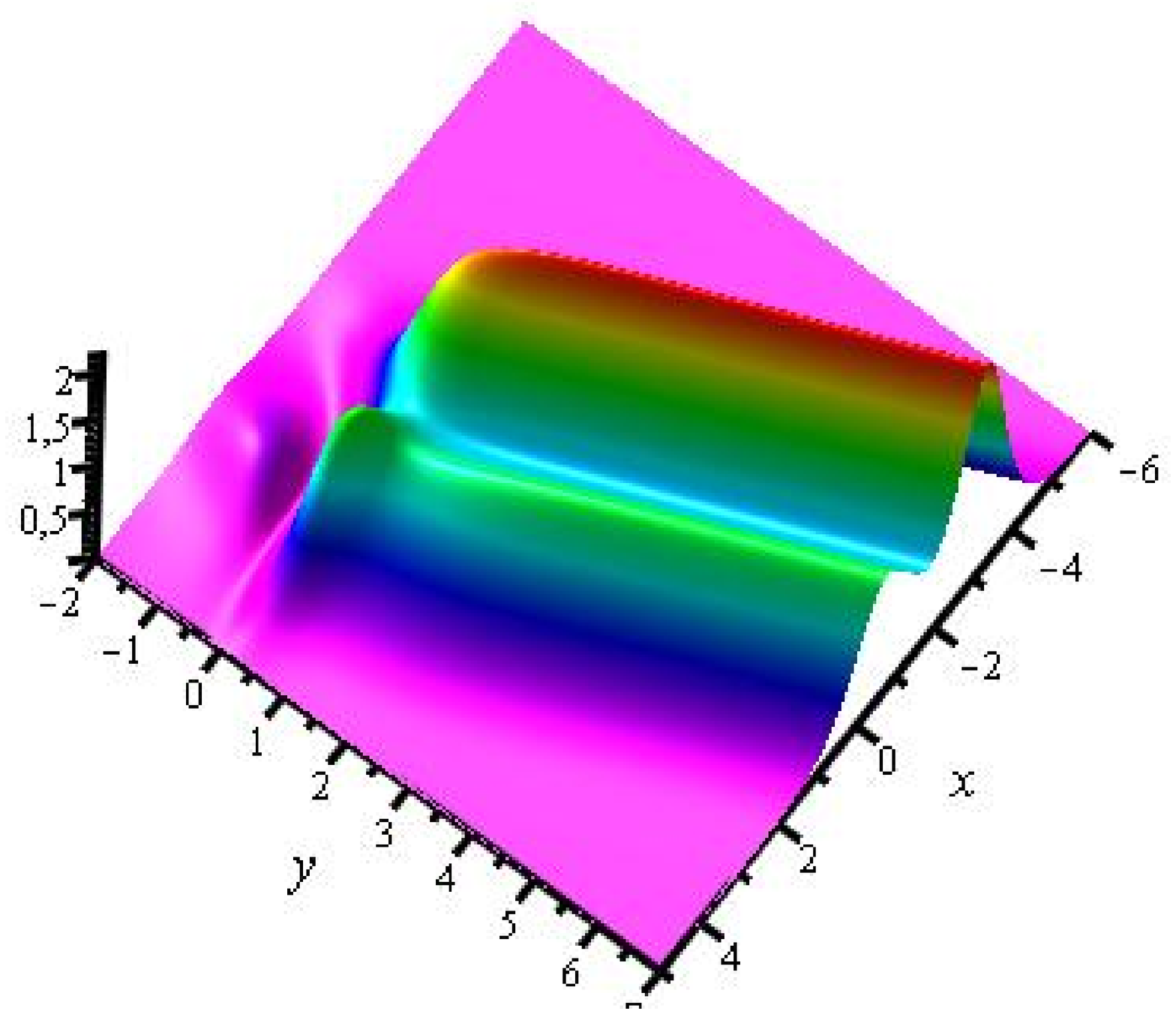}}\hspace{2cm}
\subfloat[] {\includegraphics[width=0.35\textwidth]{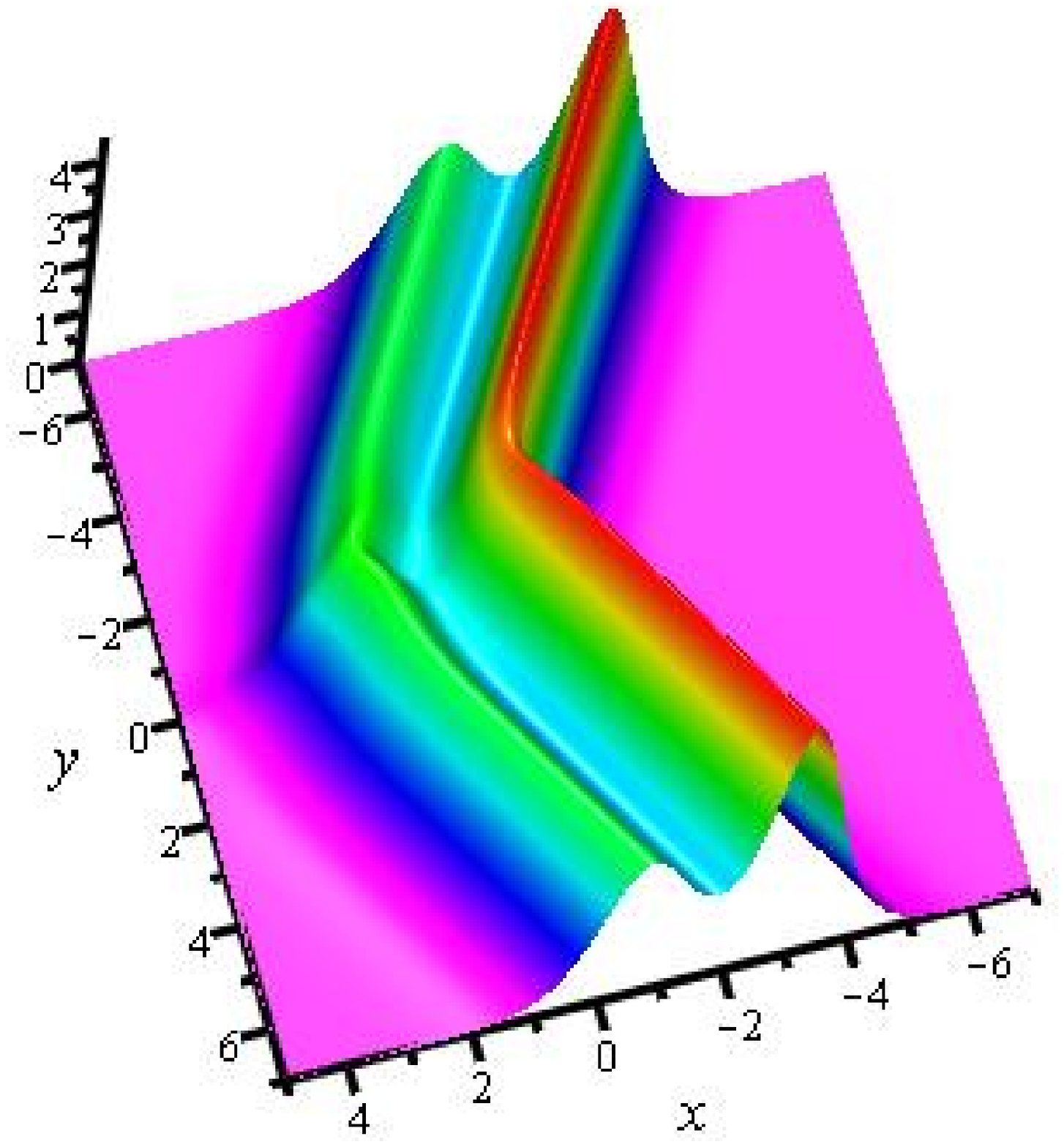}}
\caption{Two-soliton solutions of the nonlocal system (\ref{realnon2-a}) and (\ref{realnon2-b}) at $t=0$ for the parameters  $(k_1, k_2, l_1,l_2,a)=(1,2,\frac{1}{2},1,i)$
 with $\sigma_1=e^{\delta_1}=e^{\delta_2}=\rho=1$. (a) Localized combined two solitoffs from $|u(x,y,t)|^2$, (b) Localized combined two V-type solitary waves from $p(x,y,t)$.}
\end{figure}
\end{center}
\squeezeup

In Figure 13(a) there occurs a bump near the combined solitoffs. The amplitudes of the solitoffs change as time changes but they are localized. This property is also same for combined V-type waves in Figure 13(b). Figures 14(a) and 14(b) show the interactions of the waves from $|u(x,y,t)|^2$ and $p(x,y,t)$ for different times in different positions at the $y$-axis, respectively.
\begin{center}
\begin{figure}[h!]
\centering
\subfloat[]{\includegraphics[width=0.29\textwidth]{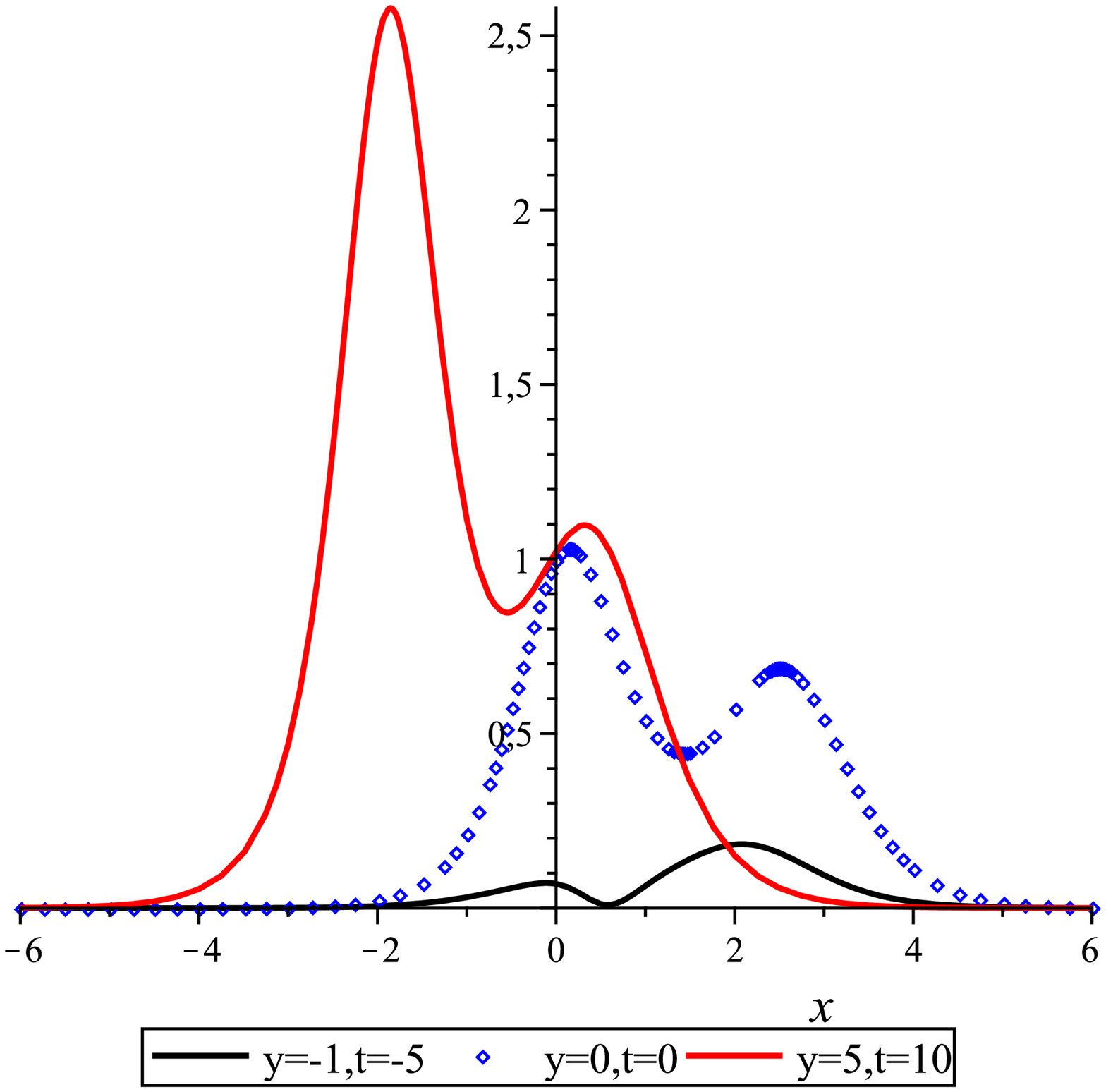}}\hspace{2cm}
\subfloat[] {\includegraphics[width=0.29\textwidth]{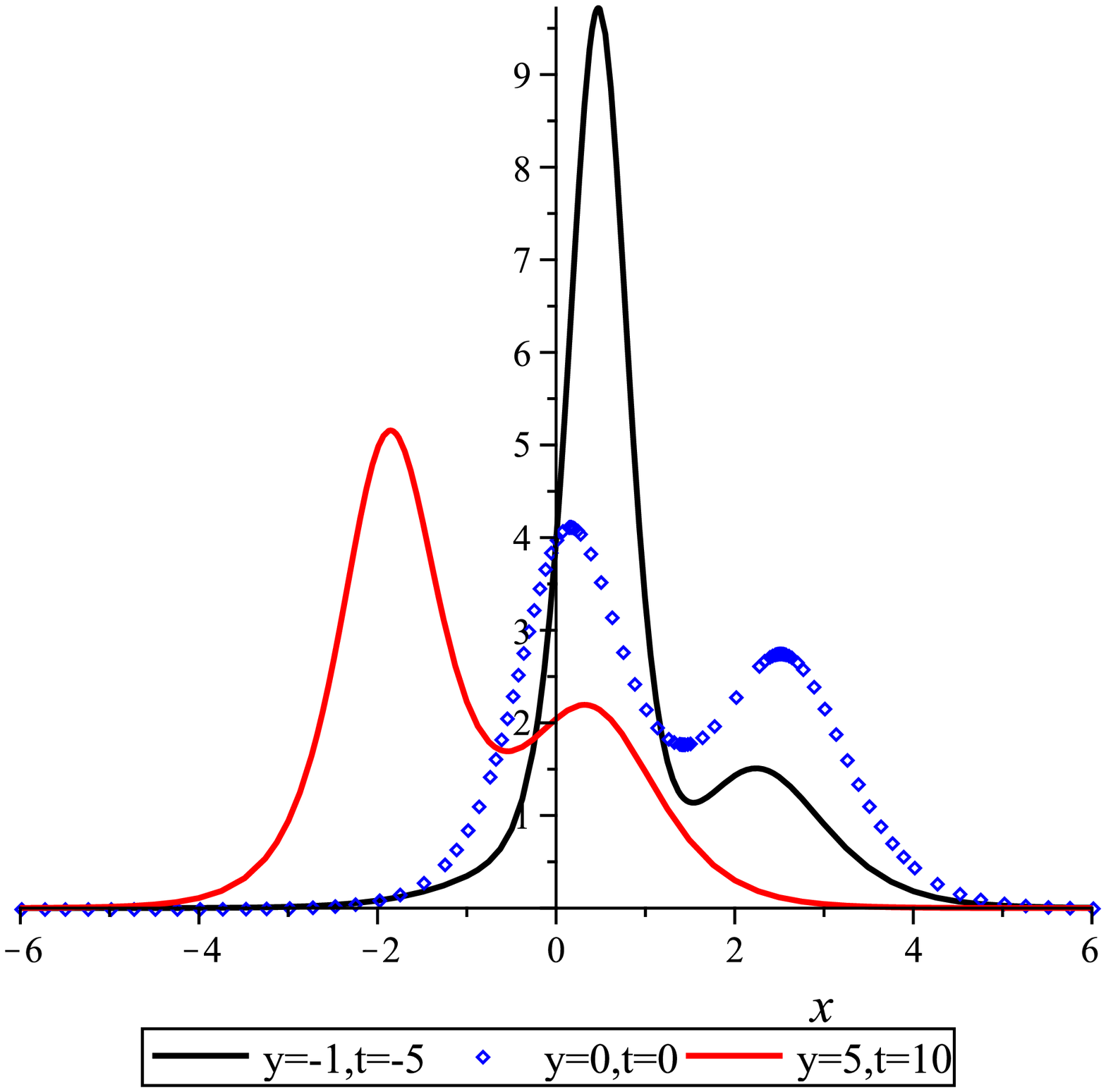}}
\caption{The movement of the waves as time increases.}
\end{figure}
\end{center}
\squeezeup
\noindent \textbf{(b).(iii)} $(\varepsilon_1,\varepsilon_2,\varepsilon_3)=(-1,-1,-1)$, $\sigma_2=\sigma_1$, $\sigma_1=\pm 1$.

In this case Type 1 gives trivial solution. Therefore we apply Type 2 approach and obtain the constraints,
\begin{equation}
  s_j=l_j,\quad e^{\alpha_j}=\sigma_3 e^{\delta_j},\quad j=1, 2,
\end{equation}
and
\begin{equation}
e^{2\delta_1}=\frac{\rho\sigma_1(k_1+\bar{k}_2)(k_1+\bar{k}_1)(l_1+\bar{l}_2)(l_1+\bar{l}_1)}{
  \sigma_3(l_1-l_2)(k_1-k_2)} ,\quad e^{2\delta_2}=\frac{\rho\sigma_1(k_2+\bar{k}_1)(k_2+\bar{k}_2)(l_2+\bar{l}_1)(l_2+\bar{l}_2)}{
  \sigma_3(l_1-l_2)(k_1-k_2)},
\end{equation}
where $\sigma_3=\pm 1$. Nonsingular two-soliton solutions obtained here are very similar to the ones given in (b).(i) case. Therefore we
will not give a particular example corresponding to this case.

\noindent \textbf{(b).(iv)} $(\varepsilon_1,\varepsilon_2,\varepsilon_3)=(-1,1,-1)$, $\sigma_2=-\sigma_1$, $\sigma_1=\pm 1$.

In this case using both Type 1 and Type 2 approaches gives trivial solutions $u(x,y,t)=0$ and $p(x,y,t)=0$.\\

\section{Conclusion}
In this work we considered integrable $(2+1)$-dimensional $3$-component Maccari system which has many applications in several areas such as hydrodynamics, plasma physics, Bose-Einstein condensates, and nonlinear optics. Under some special transformations, this system can be reduced to well-known systems like NLS, Fokas, and long wave resonance systems. We obtained one- and two-soliton solutions of the $3$-component Maccari system by using the Hirota bilinear method. Then we presented all possible local and nonlocal reductions of the system and obtained several new integrable systems. By using one- and two-soliton solutions of the $3$-component Maccari system with reduction formulas, we obtained soliton solutions of the reduced integrable $2$-component Maccari systems. Types of the one-soliton solutions obtained for these reduced local and nonlocal systems are solitons, solitoffs, and V-type solitary waves. Two-soliton solutions of reduced $2$-component systems define interactions of mentioned type of waves.

As a future work we plan to obtain different types of exact solutions to local and nonlocal reductions of $3$-component Maccari system. In recent years, rogue wave (also known as freak wave, killer wave) phenomenon has become a hot topic in the nonlinear science. Rogue waves are large and spontaneous ocean surface waves. They also appear in optics, superfluids, Bose-Einstein condensates, cold matter systems, and so on.  By finding
higher-order soliton solutions via the Hirota bilinear method and applying long wave limit we can obtain rogue wave solutions \cite{RW1}-\cite{RW5} to the $3$-component Maccari system and its local and nonlocal reductions. To study integrability properties of the Maccari system we will consider Lie symmetry analysis of this system \cite{Kumar1}-\cite{Kumar5}. Under this analysis we can also obtain new solutions of the $3$-component Maccari system and its reductions.


\end{document}